\documentclass[useAMS,usenatbib]{mn2e}

\usepackage{natbib}
\usepackage{graphicx, subfigure}
\usepackage{amssymb,amsmath}

\def \picWidth {56mm}

\title[MOCCA code for star cluster simulations - I. Blue Stragglers, first results]{MOCCA code for star cluster simulations - I. Blue Stragglers, first results}
\author[A. Hypki and M. Giersz]{
	Arkadiusz Hypki$^{1}$\thanks{E-mail: ahypki@camk.edu.pl} and 
	Mirek Giersz$^{1}$\\
$^{1}$N. Copernicus Astronomical Center of the Polish Academy of Sciences, ul. Bartycka 18, 00-716, Warsaw, Poland
}

\begin{document}

\date{Accepted YYYY Month DD. Received YYYY Month DD; in original form YYYY Month DD}

\pagerange{\pageref{firstpage}--\pageref{lastpage}} \pubyear{YYYY}

\maketitle

\label{firstpage}
 
\begin{abstract}

We introduce an improved code for simulations of star clusters, called MOCCA. It combines the Monte Carlo method for star cluster evolution and the \textit{Fewbody} code to perform scattering experiments. The \textit{Fewbody} was added in order to track more precisely dynamical interactions between objects which can lead to creations of various exotic objects observed in the star clusters, like Blue Stragglers Stars (BSS). The MOCCA code is currently one of the most advanced codes for simulating real size star clusters. It follows the star cluster evolution closely to N-body codes but is much faster. We show that the MOCCA code is able to follow the evolution of BSS with details. It is a suitable tool to perform full scale evolution of real star clusters and detail comparison with observations of exotic star cluster objects like BSS.

This paper is the first one of the series of papers about properties of BSS in star clusters. This type of stars is particularly interesting today, because by studying them one can get important constrains on a link between the stellar and dynamical evolution of star clusters. We discuss here first results concerning BSS for an arbitrary chosen test model. We investigate properties of BSS which characterize different channels of formation like masses, semi-major axes, eccentricities, and orbital periods. We show how BSS from different channels change their types, and discuss initial and final positions of BSS, their bimodal distribution in the star cluster, lifetimes and more.

\end{abstract}

\begin{keywords}

stellar dynamics - methods: numerical - globular clusters: evolution - stars: Blue Stragglers

\end{keywords}

\section{Introduction}
\label{sec:Introduction}

This is the first paper of the series of papers about simulations of star clusters, using improved version of the Monte Carlo code, called MOCCA, which stands for MOnte Carlo Cluster simulAtor. The MOCCA code is currently one of the most advanced codes which is able to simulate real size star clusters and at the same time, it allows to have a full dynamical history of the evolution of all stars in the system. It follows the star cluster evolution closely to N-body codes but is much faster. We show that the MOCCA code is able to follow the evolution of exotic objects, in this case Blue Stragglers Stars (BSS), with details and thus it is a suitable tool to perform meaningful analysis and comparisons with observations of exotic star cluster objects. \citet{Chatterjee2010ApJ...719..915C} are among the first who showed that the Monte Carlo method can be a very good tool for studying BSS.

The main subject of this paper concerns first results about statistics of BSS. They are particularly interesting today, because by studying these types of objects, one can get important constrains on a link between stellar and dynamical evolution of the star clusters. Star clusters are very efficient environments for creating exotic objects like BSS. By studying them one can reveal the dynamical history of a cluster and the role of dynamics on the stellar evolution. BSS statistics can also provide some constrains for initial binary properties. Thus, numerical codes which are able to simulate the complete evolution of star clusters and at the same time follow movements, dynamical interactions and evolution of any stellar objects, are very significant. Results of such simulations, by comparing with observational data, can verify many theories. 

BSS are defined as stars which are brighter and bluer (hotter) than the main sequence turn-off point. These stars lie along an extension of the main sequence (MS) in the Color-Magnitude Diagram (CMD) and appear to be rejuvenated stellar population. BSS are on the place in CMD where they should already evolve away from the main sequence. Their mass is larger that the turn-off mass, which suggests some stellar merger or a mass transfer scenario for their creation. They were first discovered by \citet{1953AJ.....58...61S} in M3 and later observations showed that BSS are present essentially in all star clusters. \citep{2004ApJ...604L.109P} counted 3000 BSS in 56 different size clusters. 

Currently, there are two main scenarios considered as possible formation mechanisms for BSS. The first scenario is a mass transfer between binary companions which can lead to the coalescence of the binary system \citep{1964MNRAS.128..147M,1976ApJ...209..734Z}. The second leading scenario for creating BSS is a physical collision between stars \citep{1976ApL....17...87H}. However, the exact nature of channels of formation of these objects is still unclear. According to \citet{1992AJ....104.1831F}, different environments could be responsible for different origins of BSS. In globular clusters (GCs) which are not dense, BSS could form as evolutionary mergers of primordial binaries, and in high density GCs, BSS could form from dynamical interactions, particularly from interactions involving binaries.

Relative efficiency of these two main formation channels is still unknown. Though, it is believed, that they act with different efficiency according to the cluster structural parameters \citep{1992AJ....104.1831F} and additionally they can work simultaneously in different radial parts of a star cluster \citep{1997A&A...324..915F, 2006MNRAS.373..361M}. Particularly, the number of BSS formed in the cluster does not correlate with the predicted collision rate \citep{2004ApJ...604L.109P, 2008ApJ...678..564L, 2008IAUS..246..331L}. This is one of the reasons why it is believed that mass transfer mechanism is more important in creation of BSS instead of collision between stars \citep{Knigge2009Natur.457..288K}.
Unfortunately, there is still no simple observational distinction between BSS formation through mass transfer or collision between stars. One of the first attempts to clarify this issue is the approach of \citet{2009RMxAC..37...62F}, who observed a significant depletion of C and O suggesting mass transfer mechanism for creating some BSS sub-population in 47~Tuc.
According to \citet{Davies2004MNRAS.349..129D} primordial binaries with BSS are vulnerable to exchange encounters in the crowded environments of star clusters. Low-mass components are replaced by more massive single stars. The authors claim that these encounters tend to reduce the number of binaries containing primaries with masses close to the present turn-off mass. Thus, the population of primordial BSS is reduced in more massive star clusters.

\citet{2003ApJ...588..464F} defined the BSS specific frequency as the number of BSS, normalized to the number of the horizontal branch stars. They examined 6 GCs and found that BSS specific frequency varies from 0.07 to 0.92, and it does not depend on central density, total mass and velocity dispersion. What is surprising, they found the largest BSS specific frequencies for clusters with the lowest central density (NGC~288) and the highest central density (M80). \citet{2003ApJ...588..464F} claim that these two kinds of BSS formation processes, mass transfer and mergers, can have comparable efficiency in producing BSS in their respective typical environments.
\citet{2008A&A...481..701S} found a strong correlation between BSS specific frequency and linear combination of binary fraction ($\xi_{bin}$) and velocity dispersion ($\sigma_{v}$) $\xi_{bin} + \alpha \sigma_{v}$ where $\alpha = -4.62$. This indicates that, for a given binary fraction, BSS specific frequency decreases with increasing velocity dispersion. Small cluster velocity dispersion corresponds to a lower binding energy limit between soft and hard binaries (to a larger fraction of hard binaries). Since the natural evolution of hard binaries leads to increase of their binding energy \citep{Heggie1975MNRAS.173..729H}, low velocity dispersion GCs should host a larger fraction of hard binaries, which are able to both survive possible stellar encounters, and activate mass-transfer and/or merging processes between the companions \citep{2008A&A...481..701S}. Therefore, more BSS formed by the evolution of primordial binaries are expected to form in lower velocity dispersion GCs. 

\citet{2008A&A...481..701S} tested 13 low-density GCs for correlations between specific frequency of BSS and cluster parameters like binary fraction, total magnitude, age, central velocity dispersion, metallicity, cluster central density, half-mass relaxation time, half-mass radius, stellar collision rate, concentration, and cluster evaporation rate. BSS specific frequency was defined as ratio between estimated BSS number and MS number. MS were chosen, instead of horizontal branch (HB) or red giant branch (RGB) stars, because of their abundance in all clusters and its completeness. They found the strongest correlation between number of BSS and binary fraction. It suggests that the primordial binaries fraction is one of the most important factor for producing BSS. Additionally, noticeable correlation exists with the absolute magnitude and anticorrelation with the cluster age and central velocity dispersion. The age estimates are uncertain and span a narrow range, so one has to be careful while making some generalizations. However, if such anticorrelation with the cluster ages would be confirmed in the future, it could suggest that binaries disruptions in cores of GCs become more efficient with time, which would in consequence reduce the fraction of binaries and also BSS in the core. \citet{2008A&A...481..701S} suggest that strong correlation between number of BSS and binary fraction is a result of formation channel of BSS as the unperturbed evolution of primordial binary systems. They found no correlations for central density, concentration, stellar collision rate and half-mass relaxation time. This indicates that the collisional channel of BSS formation has very small efficiency in low-density GCs.

Radial distribution of BSS in many clusters is bimodal. First discoveries of bimodal distributions were done for the M3 by \citet{Ferraro1993AJ....106.2324F, 1997A&A...324..915F} and by \citet{Zaggia1997A&A...327.1004Z} for M55. BSS radial distribution for M3 cluster clearly shows the maximum at the center of the cluster, clear-cut dip in the intermediate region and again rise of BSS in the outer region of the cluster (but lower than the central value). Bimodal distribution of BSS was later shown by other authors for other clusters like 47~Tuc \citep{2004ApJ...603..127F}, NGC~6752 \citep{2004ApJ...617.1296S}, M55 \citep{2007ApJ...670.1065L}, M5 \citep{2006ApJ...646..881W, 2007ApJ...663..267L}, and others. \citet{1994ApJ...431L.115S} suggested that BSS were formed by direct collisions in the center of the cluster and then ejected to the outer part of the system as a result of a dynamical interaction. Ejected BSS would afterwards moved back to the center of the cluster because of the mass segregation, which leads to the increase of the number of BSS in the center of the system. If the dynamical interaction is energetic enough then BSS would stay outside of the cluster for longer time and this could be the reason why there is a higher rate of BSS in the outer part of the cluster - second peak of BSS in bimodal distribution. Later, bimodal distribution of BSS in the cluster was explained differently by \citet{1997A&A...324..915F}. They showed it is a result of different processes forming BSS in different parts of the cluster - mass transfer for the outer BSS and stellar collisions leading to mergers for BSS in the center of the cluster. Furthermore, \citet{2004ApJ...605L..29M, 2006MNRAS.373..361M} and \citet{2007ApJ...663..267L} by performing numerical simulations showed that bimodal distribution in the cluster cannot be explained only by collisional scenario in which BSS are created in the center of the cluster and some of them are ejected to the outer part of the system. This process is believed to be not efficient enough, and $\sim 20-40\%$ of BSS have to be created in the peripherals in order to get the required number of BSS for the cluster. It is believed, that in the outer part of a star cluster, binaries can start mass transfer in isolation without suffering from energetic dynamical interactions with field stars. Even if one can observe bimodal distribution of BSS for many clusters, one can not generalize this feature. There are known clusters for which radial distributions are even flat, like for NGC~2419 \citep{2008ApJ...677.1069D,Contreras2012ApJ...748...91C}. 

\citet{Ferraro2006ApJ...647L..53F} gave first results of currently being performed extensive surveys of chemical composition of BSS for some selected GCs. They examined 43 BSS in 47~Tuc and found the first evidence that some sub-population of these BSS have significant depletion of C and O with respect to the normal cluster stars. They argue that this is caused by CNO burning products on BSS surface, coming from the core of a deeply peeled primary star. This scenario is expected for the case of mass transfer formation mechanism and could be the first direct proof of this formation process. Later, \citet{Fossati2010A&A...510A...8F} attempted to develop a formation scenario for HD~73666, a known BSS from the Praesepe cluster, and showed that abundance of CNO is consistent with a collisional formation. 
However, they were unable to determine whether HD~73666 is a product of a collision between two stars, components of a binary or between binary systems. Further studies of these phenomena could create some statistics on how efficient this mechanism could be in producing BSS.

\citet{Ferraro2009Natur.462.1028F} reported two distinct sequences of BSS in the globular cluster M30. These two groups are clearly separate in the CMD and nearly parallel to each other \citep[Fig. 1]{Ferraro2009Natur.462.1028F}. The first BSS sequence was accurately reproduced by the collisional isochrones \citep[Fig. 4, blue points]{Ferraro2009Natur.462.1028F}. The second BSS sequence well corresponds to the ZAMS shifted by 0.75 mag, marking the position of the low-luminosity boundary predicted for a population of mass-transfer binary systems \citep[Fig. 4, red points]{Ferraro2009Natur.462.1028F}.

\citet{Knigge2009Natur.457..288K} focused on BSS in cores of star clusters, because in these regions collisions between stars should be the most frequent. They used existing data from a large set of HST-based CMDs and confirmed that there is no global correlation between the observed core BSS number and the collision rate (different core densities have different predicted collision rates and it does not correlate with the number of BSS). However, there is a significant correlation if one would restrict this relation to the clusters with dense cores (see \citet{Knigge2009Natur.457..288K} black points in Fig. 1). The second relation which was tested by this group concerns the binary fraction in the core. If most of BSS are formed in binaries, the number of BSS should scale with the binary fraction simply as $N_{BSS} \propto f_{bin} M_{core}$, where $f_{bin}$ is the binary fraction in the core, and $M_{core}$ is the total stellar mass contained in the core. Indeed, they found clear correlation between the number of BSS and core masses of the clusters, as it is expected for the scenario where most BSS originate from binaries (see \citet[Fig.~2]{Knigge2009Natur.457..288K}). They interpret this result as a strong evidence that more BSS originates from binaries instead of collisions between stars. They found that dependence $N_{BSS} \propto M_{core}^{\delta}$ can be estimated with $\delta \simeq 0.4-0.5$. Furthermore, they estimated power law correlation $f_{bin} \propto M_{core}^{-0.35}$ based on the data from \citet{2008MmSAI..79..623M} who described global parameters for 35 clusters spanning a wide range of density and other dynamical star cluster parameters. Those two estimates combined together shows that the number of BSS found in the cores of globular clusters scales roughly as $N_{BSS} \propto f_{bin} M_{core}$, just as expected if most core BSS are formed in binary systems \citep{Knigge2009Natur.457..288K}.

BSS are being found in the halo and in the bulge of the Galaxy \citep{Bragaglia2005IAUS..228..243B,Fuhrmann2011MNRAS.416..391F,Clarkson2011ApJ...735...37C}. \citet{Tillich2010A&A...517A..36T} found that a star SDSSJ130005.62+042201.6 (J1300+0422 for short) is a BSS from the halo and has radial velocity of about $504.6 \pm 5$~km/s. With Galactic rest-frame velocity of about 467 km/s, J1300+0422 travels faster than any known blue straggler, but still is bound to the Galaxy.

Recently, \citet{Geller2011Natur.478..356G} reported that BSS in long period binaries in an old (7 Gyr) open cluster, NGC~188, have companions with masses of about half of the solar mass, which is a surprisingly narrow mass distribution. This rules out a collisional origin for these long period BSS, because otherwise, for collision hypothesis there would be significantly more companions with higher masses. The data is consistent with a mass transfer origin for the long-period blue straggler binaries in NGC 188, in which the companions would be white dwarfs of about half of a solar mass \citep{Geller2011Natur.478..356G}. 

At the end of this section we would like to note that in the literature, terms \textit{collision} and \textit{merger} are used differently by different authors. In this work the term \textit{collision} is defined as a physical collision between at least two stars during a dynamical interaction, while the term \textit{merger} is defined as a coalescence between stars from one binary as a result of stellar evolution.

This paper is organized as follows. In the Sec.~\ref{sec:InterfaceToFewbody} there is described the MOCCA code, its advantages and drawbacks in comparison to N-body codes, its design and interface between different internal parts of the code. In the Sec.~\ref{sec:Simulations} there is a description of the initial conditions for the test simulation, used in this paper, which shows the ability of the code to follow the evolution of exotic objects (BSS in this case). Sec.~\ref{sec:ChannelsOfFormationForBSS} contains first results of the simulation and physical interpretation of properties of BSS for different channels of formation. We will discuss BSS masses, orbital periods, eccentricities, BSS type changes and possible induced mass transfer which could lead to BSS formation. Moreover, we will discuss BSS global properties, like their initial and final positions in the star cluster, BSS bimodal distribution and their lifetimes. Final Sec.~\ref{sec:Summary} summarizes our findings, presents discussion about channels of formation of BSS, and highlights the future plans for the MOCCA code.

\section{Description of the code}
\label{sec:InterfaceToFewbody}

We give here a description of internal parts of the MOCCA code and the interface which connects them into one package. In the first subsection there is described the old version of the code, which is basically Monte Carlo method for star clusters simulations together with dynamical interactions based on cross-sections, and code for stellar evolution. Then, we describe the \textit{Fewbody} code \citep{Fewbody2004-01-004}, which was added to deal with the dynamical interactions between binaries and single stars or between two binaries. Next, we give the technical reference of the interface which connects these two codes. In the last subsection, there is a brief summary about the performance of the code.

This paper introduces for the first time the MOCCA code. Thus, in this section there is a detailed technical description of its internal parts, particularly the interface which connects them. This description is important to understand the design of the code and to be able to develop new procedures into the MOCCA code.

\subsection{Old version of the code}
\label{sec:IntroductionToMOCCA}

The old version of the code refers to the code which uses Monte Carlo approach to describe the relaxation processes for star clusters but with some additional features which have been developed over the years by Giersz and his collaborators \citep{Giersz1998MNRAS.298.1239G, Giersz2001MNRAS.324..218G, Giersz2006MNRAS.371..484G, Giersz2008MNRAS.388..429G}. The code for Monte Carlo method itself is in turn based on the orbit-averaged Monte Carlo method developed in the early seventies by \citet{Henon1971Ap&SS..14..151H} and then substantially improved by \citet{Stodolkiewicz1986AcA....36...19S}. Additionally, the old version of the code is equipped with procedures for performing three- and four-body dynamical interactions, based on a cross-section approach, involving primordial binaries and binaries formed dynamically. It takes into account the Galactic tide during escape processes, according to the theory proposed by \citet{Baumgardt2001MNRAS.325.1323B}, and it has implemented procedures to perform the internal stellar evolution of both single and binary stars \citep{Hurley2000MNRAS.315..543H, Hurley2002MNRAS.329..897H}. Details about the old version of the code one can find in \citep[and references within]{Giersz2011arXiv1112.6246G}. 

The old version of the code had several shortcomings. Dynamical interactions between objects were calculated analytically, using cross-sections. The problem is that cross-sections are not well known in the case of unequal masses and complicated resonant interactions. Additionally, for cross-sections possibility of stellar collisions are excluded. This is one of the most important shortcomings of the old version of the Monte Carlo code. The \textit{Fewbody} code fixes this problem and in the current version of the MOCCA code there are possible all kinds of outcomes for dynamical interactions. Moreover, the \textit{Fewbody} allows to have binaries hardening and softening with respect to only hardening in the old version of the code. This is a very important improvement which opens a lot of new interaction channels and allows to precisely track changes of energy of binaries. Details about the advantages of using the \textit{Fewbody} one can find in the Sec.~\ref{sec:DescriptionOfFewbodyCode}.

The second main shortcoming of the old version of the Monte Carlo code concerned escaping stars from a tidal limited star cluster. The escape rate was scaling with $N^{3/4}$ to deal with backscattering process \citep{Baumgardt2001MNRAS.325.1323B} but a star or a binary was removed from a star cluster immediately. To describe more realistically the escape process we implemented procedures based on \citet{Fukushige2000MNRAS.318..753F}. The escape is not anymore immediate but stars need some time to escape from a star cluster. It can significantly delay the time of the escape of a star from the star cluster. Details one can find in the next paper of this series \citep{Giersz2011arXiv1112.6246G}.

\subsection{The \textit{Fewbody} code}
\label{sec:DescriptionOfFewbodyCode}

The \textit{Fewbody} \citep{Fewbody2004-01-004} is a software package for performing small-N scattering experiments. It uses the $8$-th order Runge-Kutta Prince-Dormand integrator to advance the particles' positions. There is a possibility to enable the full pairwise K-S regularization in the simulation too \citep{Aarseth1974CeMec..10..185A}. The \textit{Fewbody} code detects stable hierarchical systems and isolates unperturbed hierarchies to increase dramatically overall performance. Hierarchies and internal data structures of stars are stored in binary trees which means that each bound object can have only two child objects (the simplest hierarchy is a binary star). The \textit{Fewbody} uses \citet{Mardling2001MNRAS.321..398M} stability criterion to assess the stability of hierarchies at each level and interrupts calculations if all bound objects are considered as stable. This code can handle dynamical interactions between any arbitrary number of stars and understands arbitrary complicated hierarchies. Full details about the \textit{Fewbody} code one can find in \citet{Fewbody2004-01-004}.

The \textit{Fewbody}, in the present version of the MOCCA code, is used to perform binary-single and binary-binary interactions. Incorporating it for performing scattering dynamical interactions, brings additional desired features. The MOCCA code runs independently small N-body scattering experiments like binary-binary interactions, using the \textit{Fewbody}, and afterwards results are written back to the MOCCA data structures. 

The \textit{Fewbody} in the MOCCA code allows to follow the dynamical evolution of binary objects and higher hierarchies just like in N-body simulations, but still giving opportunity to run simulations several orders of magnitudes faster than the N-body codes (see Sec.~\ref{sec:CodePerformance}). The \textit{Fewbody} allows the MOCCA code to perform interactions between stars and binaries more realistically than using analytical formulas like in the old version. In the beginning of the clusters simulations the most important process is the stellar evolution but after some time, in many clusters, dynamical interactions between stars start to play a huge role in the overall clusters evolution. Thus, dynamical interactions can be very significant not only for the global cluster evolution but also if one has to study creation and evolution of many different, exotic objects like compact binaries, black holes, BSS and more. Thus, the \textit{Fewbody} is so much needed to have more realistic simulations. 

The channels of formation of BSS are the main subject investigated in this paper. The \textit{Fewbody} allows the MOCCA code to have channels of formation essentially the same as in N-body simulations. It is possible to have any outcome from interactions between stars and binaries, like single or multiple mergers, exchanges and disruptions, both in simple dynamical interactions and for resonant ones. Resonant interactions occur when an incoming star is temporarily bound to the binary during the scattering experiment. Eventually one star has to escape, but before it happens, some complex and chaotic stars movements can occur. In the old version of the code it was not possible to get all complex outcomes from dynamical interactions. In comparison to the old version of the code, we expect to have similar global properties of star clusters, but parameters which strongly depend on dynamical interactions (like number of BSS or spatial distribution of BSS) should be closer to observations or results coming from pure N-body simulations.

Input parameters for the \textit{Fewbody} are stars' and binaries' parameters such as masses $m_i$, radii $R_i$, semi-major axes $a_i$, eccentricities $e_i$, and some global values which characterize a dynamical interaction like impact parameter $b$, relative velocity at infinity $v_{\infty}$, and technical parameters like tidal perturbation, maximum time for computations in seconds $t_{CPU}$, dynamical time in years $t_{dyn}$, or KS regularization -- a coordinate transformation that removes all singularities from the N-body equations, making the integration of close approaches much more accurate.

Impact parameter, $b$, is divided by a sum of the semi-major axes of all interacting binaries. Relative velocity, $v_r$, is the relative velocity at the infinity between interacting distinct bound objects in terms of the critical velocity, $v_c$, defined in a way that the total energy of the binary-single or binary-binary system is zero. If $v_{\infty} > v_c$ the total energy of the system is positive and it is possible that each object will leave the system unbound from any other with positive velocity at infinity. If $v_{\infty} < v_c$ the total energy is negative and the encounters are likely to be resonant, with all stars involved remaining in a small volume for many dynamical times. 

Tidal perturbation determines when analytic formulas and direct integration procedures are used. Each numerical integration is started at the point at which the tidal perturbation on a binary in the system reaches some specified value $\delta$ ($\delta = 10^{-5}$ is the default value in the MOCCA code). 
Tidal perturbation is defined as $F_{tid} / F_{rel}$, where $F_{tid}$ is the tidal force at the apocentre, and $F_{rel}$ is the relative force at the apocentre (for details see \citet{Fewbody2004-01-004}). The same mechanism is used internally by the \textit{Fewbody} to speed up integration between stars and bound objects. The force is not calculated between all stars but between objects which do not break this tidal perturbation threshold. It is a quite important parameter because smaller values of $\delta$ yield better energy conservation but increase the computational time -- more integration steps are calculated with numerical integrator rather than with analytical equations (for more information about the hierarchy isolation see \citet{Fewbody2004-01-004}).

Maximum time of computations in seconds for each \textit{Fewbody} scattering experiment is set by parameter $t_{CPU}$. After this time the interaction is forced to stop. It is possible that stars are still close to each other (tidal perturbation is still $> \delta$). Thus, this parameter has to be chosen carefully. It can not be too short because many more dynamical interactions would not be completed according to stopping conditions (described later in this section). By experimenting and calculating how many interactions were not completed, $t_{CPU} = 10$~s was chosen as an optimal value. For maximum time 10 seconds for a simulation with 100k objects there were only few dozens of not completed encounters (from total 4057 interactions) -- it seems to be a reasonable value. It is also worth to mention, that it is hard to predict in advance how many interaction steps each scattering experiment would take, so a better solution is to use $t_{CPU}$. Interrupted interactions resulted in creating unstable or stable triples or quadruples. These objects were artificially disrupted to binaries and single stars because the Monte Carlo part of the MOCCA code is currently unable to handle complex hierarchies. However, even if those objects were manually disrupted, the binding energy of triples and quadruples were insignificant in comparison to the average binding energy of binaries in the system. Thus manual disruption most probably had no significant influence on the overall cluster simulation. 

KS regularization for our simulations is disabled by default. Regularization transforms coordinates of stars removing all singularities from N-body equations. It allows to integrate close approaches and even collision orbits much more accurate. But using the KS regularization requires additional efforts to detect physical collisions, because pericenter is not necessarily resolved by the integrator. With enabled KS regularization there were only a few mergers per simulation. Thus, we decided to switch it off completely. For more information about the \textit{Fewbody} input parameters see \citet{Fewbody2004-01-004}.

When describing the \textit{Fewbody} one has to realize what are the \textit{Fewbody} stopping conditions, because it is crucial to find optimal parameters for the MOCCA code to have a good error conservation with decrease in performance as low as possible. Better energy conservation is for smaller $\delta$, but the \textit{Fewbody} dynamical interactions are calculated significantly longer. The \textit{Fewbody} uses several criteria to automatically terminate the integration of the scattering encounter. In general calculations are interrupted when there is no chance to bound objects to interact with each other, and bound object will not evolve internally. Stopping conditions are described in details in the \citet{Fewbody2004-01-004}.

\subsection{Interface design}
\label{sec:InterfaceDesign}

Writing an interface between Monte Carlo code and the \textit{Fewbody} code was a very complicated task. These codes have different data types storing data, different programming styles, and what is the most important, are written in different programming languages. The \textit{Fewbody} is written in C language and the Monte Carlo code is written in Fortran77. 

The approach to describe tree structures in the code, without actually having structures in Fortran77, was to use the same idea as in relational models for database management. It is a database model based on the first-order predicate logic, first formulated and proposed by \citet{DBLP:persons/Codd69}. Each table describes one data type, in other words, one entity with some finite set of attributes. Each attribute corresponds to a column in such a database and each data type corresponds to a table. There are many advantages by organizing data in such a way. Data types are easy to extend in both directions -- higher hierarchies (triples, quadruples etc.) and more attributes describing objects (like mass, radius, luminosity for stars, or semi-major axis, eccentricity for binaries). If one has to add some more attributes or more objects for an encounter, there is no need to change function declarations, but only extend what is already written to deal with new attributes. Especially, there is no need to change function declarations by adding or removing input parameters. Using such data organization one can have easy and transparent system to exchange data between very different codes. It is also worth to notice that this is just a concept of storing data. It does not involve using any unusual Fortran77 language statements. The interface simply uses plain arrays of double precision and integer numbers to store data, runs the \textit{Fewbody}, and rewrites results back to Monte Carlo structures. From now on, the term 'table' means actually just simple 2-dimensional array. 

There are defined only 3 tables to store data: \textit{hierarchy}, \textit{binary}, and \textit{single}. This is enough to describe any complex hierarchical system with any number of stars in it. 

Detailed description of these tables will be presented by the following example. 
Let us consider an interaction between two binary stars. Let us assume that the result of such interaction is one binary with one merger inside and one single stars (outside the binary). 

\begin{table}
	\begin{minipage}{.47\textwidth}
		\begin{center}
		    \begin{tabular}{ | c | c | c | c | c | c | c}
		    \hline
		    \# & Data & Parent & Child1  & Child2 & New & Merger \\  
		     & & & & & parent & id \\ \hline \hline
		    1  & -1   & 5      & 0       & 0      & 7          & 0\\
		    2  & -2   & 5      & 0       & 0      & 7          & 5\\
		    3  & -3   & 6      & 0       & 0      & 7          & 5\\
		    4  & -4   & 6      & 0       & 0      & 0          & 0\\
		    5  & 1    & 0      & -1      & -2      & 0          & 0\\
		    6  & 2    & 0      & -3      & -4      & 0          & 0\\
		    7  & 3    & 0      & -1      & -5      & 0          & 0\\
		    \end{tabular}
		    
		\end{center}
		\caption{Table \textit{hierarchy} holds information about the structure of bound objects}
		\label{tab:hierarchy}
	\end{minipage}

	\begin{minipage}{.47\textwidth}
	\begin{center}
	    \begin{tabular}{ | c | c | c | c | c}
	    \hline
	    \# & a [$AU$] & e & Binary id & New binary id\\ \hline \hline
	    1  & 1.60                  & 0.46 & 119390    & 119390 \\ 
	    2  & 0.98                  & 0.31 & 825260    & 0 \\
	    3  & 1.80                  & 0.95 & 0         & 119390 \\
	    \end{tabular}
	\end{center}
	\caption{Table \textit{binary} holds physical properties of the bound objects like semi-major axes (\textit{a}) and eccentricities (\textit{e}). The simplest bound object is a binary. Values of semi-major axes and eccentricities are exemplary.}
	\label{tab:centerofmassExample}
	\end{minipage}

	\begin{minipage}{.47\textwidth}
	\begin{center}
	    \begin{tabular}{ | c | c | c | c |}
	    \hline
	    \# & Mass [$M_{\odot}$] & Radius [$R_{\odot}$] & Star id \\ \hline \hline
	    1  & 0.76       & 0.70         & 119390 \\
	    2  & 0.75       & 0.69         & 119391 \\
	    3  & 0.98       & 0.88         & 825260 \\
	    4  & 0.60       & 0.53         & 825270 \\
	    5  & 1.74       & 4.74         & 119391 \\
	    \end{tabular}
	\end{center}
	\caption{Table \textit{single} holds information about physical properties of the stars. Masses, radii and star IDs' are exemplary.}
	\label{tab:singleExample}
	\end{minipage}
\end{table}

The data in the table \textit{hierarchy} for such encounter is shown in Tab.~\ref{tab:hierarchy}. It contains six columns: \textit{data}, \textit{parent}, \textit{child1}, \textit{child2}, \textit{new parent} and \textit{merger id}. In this table there are stored relations between stars and it contains information on which star belongs to which binary. Additionally, each row in this table describes what are the children of a specific object and which physical properties from two other tables (\textit{binary} or \textit{single}) describe this object. 

In order to describe a binary there are needed three rows in the \textit{hierarchy} table. Column \textit{data} contains row ID of the data for a single star or for a binary. In our example in Tab.~\ref{tab:hierarchy} for the two first rows we have in the column \textit{data} indices \textit{-1} and \textit{-2}. Negative values mean that these are distinct stars and their absolute values point to the first and the second row in the \textit{single} table (Tab.~\ref{tab:singleExample}). And if the column \textit{data} contains positive number then it points to the \textit{binary} table (Tab.~\ref{tab:centerofmassExample}), which contains data of binaries or higher hierarchies. Two first rows in the \textit{hierarchy} table describe both single stars of the first binary, there are no children for them, and that is why columns \textit{child1} and \textit{child2} contain \textit{0}. 

Column \textit{parent} points to the row in the same table \textit{hierarchy} and describes the parent of a given object. Both single star and binary can have parents. Of course for the simplest case, the binary-single dynamical interaction, only single stars (from a binary) can have some indices in the column \textit{parent}. Two first rows in Tab.~\ref{tab:hierarchy} contain in the column \textit{parent} values \textit{5} which corresponds to the fifth row in the \textit{hierarchy} table. It means that these two stars have the same parent and they form one binary described by the fifth row in this table. This fifth row holds information related to the binary itself. Furthermore, in the fifth row of the \textit{hierarchy} table one can see that in the column \textit{data} there is positive index \textit{1}. It points to the first row in the \textit{binary} table (Tab.~\ref{tab:centerofmassExample}) where all physical properties of such binary are stored, like semi-major axis and eccentricity.

Third and forth row in the \textit{hierarchy} table describes the first and the second star, which are bound together as a second binary. This binary containing those two stars is described by the sixth row in the \textit{hierarchy} table. One can see that in the sixth row there is the value \textit{2} which is positive and thus points to the \textit{binary} table to the second row. In this second row there are stored initial binary properties about this second binary taking part in this dynamical interaction.

The last very important column is \textit{merger id}. It points to the \textit{single} table in the case of merger. In our example there is one merger inside a binary after the interaction. One can see that in the column \textit{new parent}, which describes structure after the interaction, in the \textit{hierarchy} table there are values \textit{7} for three first rows. It means that those three single stars have the same parent. Additionally, in the column \textit{merger id} for the second and the third row, there is the value \textit{5}. It means that those two stars merged into one star in the interaction. Fifth row in the \textit{single} table describes physical properties of this merger. Thus, after the interaction the binary consists of two stars where the first star comes from the first row (\textit{new parent} column contains value \textit{7}, \textit{merger id} column is equal to \textit{0}). The second star is a merger star, produced from stars from rows \textit{2} and \textit{3} (\textit{merger id} column is equal to \textit{5}). Using these several columns one can describe any hierarchical structure needed in the MOCCA code. 

Let us now consider the last two tables (simpler ones). The table \textit{binary} (Tab.~\ref{tab:centerofmassExample}) is a table (array) which stores information about two distinct objects bound gravitationally. Such a bound object can be described using attributes like semi-major axis and eccentricity. The easiest case is a binary star, the more complicated are triples or quadruples. But in the case of triple there would be actually two binaries, one for the inner binary and the second for the outer binary, so there would be two rows in the \textit{binary} table needed to describe it.

The table \textit{single} holds properties of distinct stars (Tab.~\ref{tab:singleExample}). It contains data like masses, radii, IDs etc. Each row in this table corresponds to only one star, merger or not, involved in some binary system or not.

Rows' orders do not matter in any table. In the Tab.~\ref{tab:hierarchy} two first rows are occupied by stars belonging to the first binary, the third and forth row describe stars from the second binary. However such order is not mandatory. 

The introduced example from tables \ref{tab:hierarchy}, \ref{tab:centerofmassExample} and \ref{tab:singleExample} shows only several columns which are necessary to understand the concept behind this interface. In these tables there is stored much more information about stars and binaries. For stars there are stored additionally velocities from before and after interaction. For binaries there are stored additionally binding energies, semi-major axes and eccentricities from before and after the dynamical interaction. All these pieces of data are needed to check how properties of stars and binaries are changed after performing one \textit{Fewbody} dynamical interaction. All changed properties of the stars and binaries have to be saved to the MOCCA structures so that the Monte Carlo part of the code will be notified about these changes.

The source code of the \textit{Fewbody} is designed very well because it is stateless. It does not use any global variables (except of course for constant parameters). There was no need to worry about global variables when the interface was designed, and there was no need to change the \textit{Fewbody} code in order to incorporate it to the MOCCA code. Everything what was needed to start the \textit{Fewbody} was to run it with some initial parameters, taken from MOCCA, and then read results and propagate them into the MOCCA data structures (together with transformation to star cluster coordinates). Interface design in overall is rather complicated, however executing the \textit{Fewbody} dynamical interaction itself, is simple. The \textit{Fewbody} code design will simplify also the parallelization of the code which is planned as one of the next new features of the code (discussed later). 

Repeatable results for the simulations with exactly the same initial conditions are needed for testing purposes. Parameter, which causes that two simulations could calculate a little bit differently, is the maximum time for calculations in seconds ($t_{CPU}$). This value has further implication to the whole simulation. On different computers, during those 10 seconds, integrator in the \textit{Fewbody} can compute different number of integration steps. This will cause, that for exactly the same input parameters, the same encounter can return different results. Usually the difference concerns very small changes of such parameters like semi-major axes or output velocities. But sometimes the difference can be more significant. In the slower CPU during 10 seconds the outcome can be for example an unstable triple, but on the faster CPU, which is able to calculate more integration steps in 10 seconds, this unstable triple can be disrupted. Afterwards this small difference propagate to the whole cluster. Even a tiny difference causes that the whole simulation is not repeatable. It complicates debugging and testing. Statistically those changes have no meaning at all, the star cluster global parameters are almost the same, but it should be kept in mind that two exactly the same simulations, running even on the same machine, can give slightly different results.

\subsection{MOCCA code}

The MOCCA code is a package which combines together several other codes and allows to perform simulations of a real size star clusters. The old version was described in Sec.~\ref{sec:IntroductionToMOCCA}. Interface which combines the old version of the code with \textit{Fewbody} was described in Sec.~\ref{sec:InterfaceDesign}. All these codes together create one new package, called the MOCCA code. 

In order to combine Monte Carlo code with \textit{Fewbody} there had to be made some simplifications in the code. As was mentioned before, \textit{Fewbody} can handle any arbitrary hierarchy but not Monte Carlo. Therefore, if one of the outcomes of the \textit{Fewbody} is a hierarchical object more complicated than binary (triple or quadruple), it has to be manually disrupted. It has to be done in such a way that overall energy has to be conserved. Such cases are only count down and written to the log file. However, it is planned to adopt the MOCCA code for more complex hierarchies in the future. 

MOCCA code runs each dynamical interaction independently of any other interactions and of the host star cluster. Because of that, there are possible dynamical interactions which create very wide binaries. From \textit{Fewbody} point of view, those objects, even if they are extremely wide, are stable and they are not disrupted by the \textit{Fewbody} code. However, in star cluster one has to take into account the environment. If semi-major axes of such wide binaries exceed many times the mean distance between stars, it is most likely that these binaries would not exist at all and would be disrupted by encounters right away. Additionally, if binaries are very wide then the probability of dynamical interactions for them are very high and thus such binaries practically always take part in interactions. Furthermore, if semi-major axis is very wide, then drawn impact parameter is very large and the interaction is just a very distant fly-by (like in relaxation procedures). Those kinds of dynamical interactions are physically unimportant and occupy CPU for no reason. Thus, there were implemented procedures to disrupt those very wide binaries in binary-single dynamical interactions, according to \citet[eq. 4.12]{Heggie1975MNRAS.173..729H} probabilities formulas.

\subsection{Code performance}
\label{sec:CodePerformance}

MOCCA code is much faster than N-body codes for the same number of particles in the system. This is simply the consequence of the Monte Carlo method used in the MOCCA code. For the test model, described in details in the Sec.~\ref{sec:InitialModel}, with 120k stars, $20\%$ of primordial binaries, enabled \textit{Fewbody} for dynamical interactions, and enabled stellar evolution simulation, takes less than 4 hours to complete the calculations. Simulation with the same initial conditions but without using \textit{Fewbody} for dynamical interactions takes a little bit more than 2 hours. The computer used for simulations had AMD Opteron CPU and 4 GB of memory, so it was a simple commodity hardware. N-body codes take weeks of computation for similar models with even less initial binary fractions (e.g. \citet{Hurley2008AJ....135.2129H}). The MOCCA code is still under heavy development, it is not optimized yet and works only with one CPU. Parallelization is one of the next planned features which increase the value of the MOCCA code even more.

The speed of the MOCCA code is its great advantage in comparison to N-body codes. For the same amount of time one can run multiple simulations with the MOCCA code to cover very wide range of initial cluster parameters. Instead of having one simulation from N-body, one can have hundreds of simulations from the MOCCA code and one can perform detail statistical analysis of the results. Additionally, MOCCA simulations give practically the same amount of information about the evolution of the star clusters as N-body codes, which makes it even more attractive.

\section{Simulations}
\label{sec:Simulations}

\subsection{Initial model}
\label{sec:InitialModel}

This paper is the first one of the series of papers about properties of BSS in star clusters. There are introduced here first results concerning BSS for arbitrary chosen test model. However, in the next papers we plan to use the MOCCA code to study in details how various initial conditions of star clusters influence on the population of BSS in general and for different channels of formation. We plan to run simulations of star clusters with various initial binary fractions, cluster concentrations, metallicities, different initial number of stars and properties of primordial binaries like eccentricities and semi-major axes. However, for the need of this paper, which is a test of the MOCCA code, we chose only one model, which shows the ability of the MOCCA code to follow the evolution of BSS population in the star cluster.

As the test (initial) model we selected a model with 100k objects (80k single stars ($N_s$) $+$ 20k binaries ($N_b$)). Initial conditions used in this paper are summarized in Tab.~\ref{tab:InitialConditions}.

\begin{table}
	\begin{center}
	    \begin{tabular}{ | p{3cm} | p{4cm} }
	    Parameter  & Description \\
	    \hline
	    Single stars ($N_s$) & 80k \\
	    Binary stars ($N_b$) & 20k \\
	    Binary fraction ($f_b = \frac{N_s}{N_b + N_s}$) & 0.2\\
	    Initial model  & Plummer \\
	    IMF of stars  & \citet{Kroupa1993MNRAS.262..545K} in the range $[0.1; 50] \mathrm{M_{\odot}}$ \\
	    IMF of binaries & \citet[eq. 1]{Kroupa1991MNRAS.251..293K}, binary masses from 0.2 to 100~$\mathrm{M_{\odot}}$\\
	    Total mass (M(0)) & 6.02 $\times 10^{4} \mathrm{M_{\odot}}$ \\
	    Observational initial core radius ($r_{cl}$) & 0.36 pc \\
	    Initial half-light radius ($r_{hl}$) & 0.53 pc \\
	    Initial tidal radius & $35.8$ pc \\
	    Binary mass ratios & Uniform \\
	    Binary semi-major axes & Uniform in the logarithmic scale from $2(R_1+R_2)$ to $50$ AU\\
	    Binary eccentricities  & Thermal (modified by \citet[eq. 1]{Hurley2005MNRAS.363..293H})\\
	    Metallicity  & $0.001$ \\
	    \end{tabular}
	\end{center}
	\caption{Initial conditions for our test model}
	\label{tab:InitialConditions}
\end{table}

Masses for single stars were chosen according to the initial mass function described by \citet{Kroupa1993MNRAS.262..545K}, with following parameters:

\begin{equation}
\xi(M) = 
\begin{cases} 
M^{-1.3}, & \mbox{if } M \leq 0.5 \\ 
M^{-2.3}, & \mbox{if } M > 0.5 
\end{cases}
\label{eq:MSingle}
\end{equation} where $0.5$ is so-called brake mass. Minimum star mass was set to $0.1 \mathrm{M_{\odot}}$ and maximum to $50 \mathrm{M_{\odot}}$. All stars are assumed to be on the zero-age main sequence (ZAMS) in the beginning of the simulation. Masses of binaries are chosen from \citep[eq. 1]{Kroupa1991MNRAS.251..293K}:

\begin{equation}
M(X) = 0.33 \left [ \frac{1}{(1 - X)^{0.75} + 0.04 (1 - X)^{0.25}} - \frac{1}{1.04}(1 - X)^2 \right ]
\label{eq:MBinary}
\end{equation} where X is randomly chosen from 0 to 1. The Initial Mass Function (IMF) algorithm for binaries in MOCCA works in such a way that first it draws randomly mass for the whole binary according to the Eq. \ref{eq:MBinary}. Then, the mass ratio is drawn from a uniform distribution, and the whole mass is split into two separate masses keeping sum of the masses intact. The IMF generated a model with the total mass of 60200 $\mathrm{M_{\odot}}$ and with mass of the most massive binary of 89 $\mathrm{M_{\odot}}$. 

Tidal radius was set to 35.8 pc. Initial observational core radius was set to 0.36 pc and observational half-mass radius to 0.53 pc. Observational core radius ($r_{cl}$) is defined as a distance at which the central surface brightness drops by a half and half-light radius ($r_{hl}$) is defined as a radius within which half of the star cluster luminosity is contained. Metallicity was set to 0.001, which is typical for GCs metallicities. 

Semi-major axes were uniformly chosen from the logarithmic scale from $2(R_1+R_2)$~AU to $50$~AU, where $R_1, R_2$ are stellar radii of binaries' components. Orbital periods distribution of binaries are almost uniformly distributed, in the logarithmic scale, from $\sim 10^{0}$ up to $10^{5}$ days.

\begin{figure}
\begin{center}
  \includegraphics[width=\picWidth,angle=270]{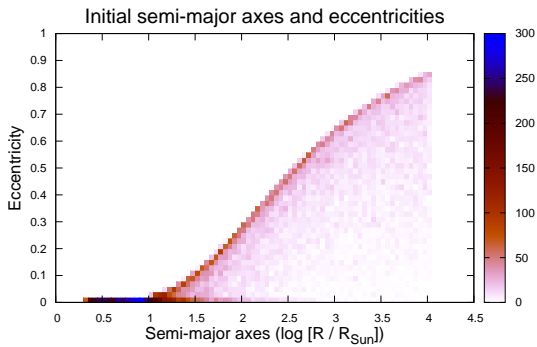}
  \caption[]{Density plot of initial semi-major axes and eccentricities. Semi-major axes are shown in the log [$ R / \mathrm{R_{\odot}}$]. The substantial fraction of binaries have eccentricities very small or equal to 0.0 and relatively small semi-major axes (around $10^{0.5} - 10^{1.0} \mathrm{R_{\odot}}$).}
  \label{fig:HistInitialConditions}
\end{center}
\end{figure}

Eccentricity ($e_b$) is chosen from thermal distribution \citep{Heggie1975MNRAS.173..729H} but with one exception. The procedure to draw eccentricities modifies eccentricity according to tidal evolution (equations 3b and 4 from \citet{Kroupa1995bMNRAS.277.1507K}). If the drawn semi-major axis is less than 5 times the ZAMS radius of the primary star, then it is assumed that a merger would happen in the beginning of the evolution, so the new set of parameters is altered, according to the following equation:

\begin{equation}
\label{eq:Eccentricity}
\mathrm{ln} e_i = - \left( \frac{\lambda \mathrm{R_{\odot}}}{R_{peri}} \right) ^{\chi} + \mathrm{ln} e_b
\end{equation}
where $\lambda = 28$, $\chi = 0.75$, and $e_i$ is the eventual eccentricity for the primordial binary. This procedure prefers small and very small eccentricities. Density plot with initial semi-major axes and eccentricities is shown in Fig.~\ref{fig:HistInitialConditions}. It shows that the substantial fraction of binaries have indeed eccentricities $\sim 0.0$ and relatively small semi-major axes (around $10^{0.5} - 10^{1.0} \mathrm{R_{\odot}}$). 

\subsection{Statistical fluctuations}
\label{sec:StatisticalFluctuations}

\begin{figure}
\begin{center} 
  \includegraphics[width=\picWidth,angle=270]{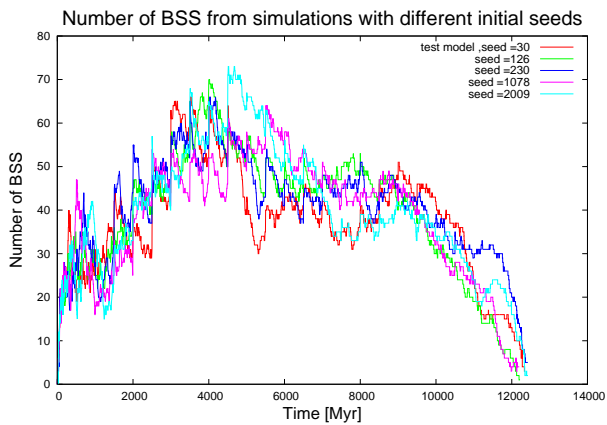}
  \caption[]{Total number of BSS from 5 simulations with the same initial conditions but with different seed values. Test model used in this paper was initialized with the seed=30.}
  \label{pic:SeedErrors}
\end{center}
\end{figure}

Fig.~\ref{pic:SeedErrors} shows the number of BSS from 5 simulations with the same initial conditions to point out how intrinsic method fluctuations influence the overall BSS population. BSS are the main subject of this paper so statistical fluctuations of their number were used here to show the strength of the fluctuations. In Fig.~\ref{pic:SeedErrors} all plots have similar characteristics. The first peak of BSS is for $\sim 1$ Gyrs, then number of BSS drops and the second main peak of number of BSS lies around 4 Gyrs. After that, it drops more or less the same for all simulations.

\begin{figure}
	\centering
		\includegraphics[width=\picWidth,angle=270]{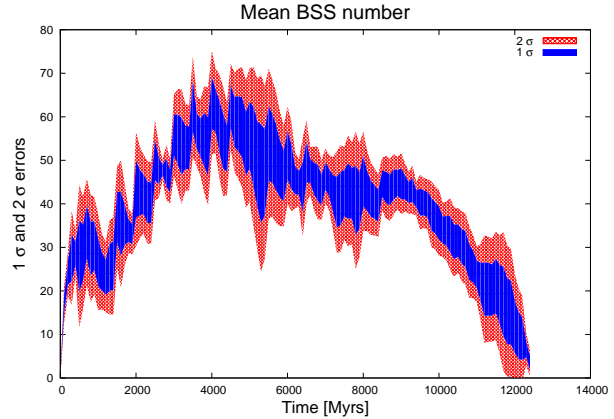}
		\caption{$1 \sigma$ and $2 \sigma$ errors for the mean number of BSS from 5 simulations with the same initial conditions but with different initial random number (seed)}
		\label{pic:SeedErrors2}
\end{figure}

Fig.~\ref{pic:SeedErrors2} shows the mean number of BSS calculated at each 100 Myrs from these 5 simulations together with $1 \sigma$ and $2 \sigma$ errors. Fig.~\ref{pic:SeedErrors} and Fig.~\ref{pic:SeedErrors2} show that different seeds do not change characteristics of BSS population significantly. 
For almost the whole simulation $1 \sigma$ is about $\pm$ 5 BSS. 
In the worst case the standard deviation from the mean value is for the time $\sim 5.3$ Gyrs (about $\pm$ 10 BSS). 

In general all features in the number of BSS which are about $\pm$ 5 BSS can be the results of the statistical fluctuations. They should not be considered in the discussion when different models are compared. General trends in number of BSS are preserved. The first peak in number of BSS ($\sim 1$ Gyrs) and the second more extended one ($\sim 4$ Gyrs) are present more or less at the same time for all 5 simulations. In the Fig.~\ref{pic:SeedErrors2} there are shown fluctuations for all BSS from all channels of formation (described in Sec.~\ref{sec:ChannelsDefinitions}), but $1 \sigma$ and $2 \sigma$ fluctuations for each channel of formation separately look very similar. 

\section{Results}
\label{sec:ChannelsOfFormationForBSS}

In the first subsection we present a description of different channels of formation of BSS, together with their physical properties. In the next subsection there is a discussion about global parameters of BSS.

\subsection{Channels of formation of BSS}
\label{sec:DefinitionsOfChannelsOfFormations}

In this chapter we introduce definitions of all channels of formation of BSS in the MOCCA code. Then, we describe their physical properties and discuss how BSS change their types during the star cluster evolution. Finally, we compare results of the MOCCA code with the old version of the code (without \textit{Fewbody}) taking into account BSS population.

All stellar evolution processes described in this paper are implemented only in single-star evolution (SSE) and binary-star evolution (BSE) part of the code \citep{Hurley2000MNRAS.315..543H, Hurley2002MNRAS.329..897H} and there are no additional evolutionary procedures implemented in the MOCCA code.

\subsubsection{Channels definitions}
\label{sec:ChannelsDefinitions}

A star is recognized as a BSS when it is a main sequence star and has a mass higher than the turn-of-mass by at least 2\% (to have the same condition as in J.~Hurley's N-body simulations and to be able to compare our simulations with N-body ones, see \citet{Giersz2011arXiv1112.6246G} for details). Similarly for observations, a star which is too close to the turn-off mass cannot be considered as BSS because of the observational errors in determination of the stellar magnitudes.

There are two main channels of BSS formation presented in the literature. The first one is a mass transfer and the second one is a collisional channel of formation. We expect that BSS created by a collision between two field stars, or between a binary and a field star, or between components of a merging binary will have different spatial distributions. The details of formation processes may also vary within one channel. Thus, instead of having just one collisional channel of formation, there will be more specific channels for distinct scenarios. In our simulations we can trace the history of each object, so it is possible to find out what was the real cause of the creation of BSS.

In this paper, for the simplicity of the discussion, we divided BSS formation channels into two general categories: stellar evolution and dynamical.

There are three channels of formation of BSS which we include to the stellar evolution category. The first one, \textit{Evolutionary Merger} (EM), describes a scenario when two stars from one binary merge into one star because of stellar evolution only. The second channel is called \textit{Evolutionary Mass Transfer} (EMT). It describes a situation when there is some mass transfer in a binary, so that the mass of one of the stars overcomes the turn-of-mass. In this case the stellar evolution does not lead to a binary merger. This scenario is described in the literature as one of the two main channels of formation -- mass transfer. The third channel is called \textit{Evolutionary Dissolution} (ED). It describes a scenario when the stellar evolution leads to a disruption of a binary (e.g. SN explosion) with some mass accretion by the companion, which in consequence becomes a BSS.

For EMT BSS it is still possible to have an evolutionary merger event after some time, but we did not observe such scenarios of changing BSS types from EMT to EM in our simulations (see later Sec.~\ref{sec:TypeChanges}). Additionally, it is worth to mention, that these evolutionary channels of formation (EM, EMT, ED) are not results of dynamical interactions, but rather of stellar evolution. It does not necessarily mean, that there were no dynamical interactions for these objects before. It only means, that BSS creation was detected after performing the stellar evolution step and it is assumed that the evolution was the direct cause of the creation of BSS. Field stars can influence binaries due to soft and rarely strong dynamical interactions. Thus, deciding whether some BSS is in fact purely evolutionary is possible only after a careful investigation of the whole history of a particular BSS.

Channels of formation of BSS which we include to dynamical category are connected strictly to dynamical interactions and are described by the following cases. The channel of formation called \textit{Collision Single-Single} (CSS) describes a physical collision between two single stars. This is the only channel, both from evolution and dynamical categories, which involves only two single stars. All other channels of formation involve at least one binary. The second channel called \textit{Collision Binary-Single/Binary} (CBS, CBB) describes the scenario when there is some collision between any two or more stars in a binary-single (CBS) or binary-binary interaction (CBB). 

The rest of the channels do not in fact create a new BSS but rather describe the change of BSS type (see Sec.~\ref{sec:TypeChanges}). \textit{Exchange Binary-Single/Binary}, corresponds to the situation when BSS changes its companion in a binary, or becomes a single star, or goes into a binary. Again, EXBS means an exchange event in a binary-single dynamical interaction and EXBB means an exchange in a binary-binary interaction. The last dynamical channel of formation is called \textit{Dissolution Binary-Single/Binary} and corresponds to the scenario when BSS was present in a binary, which was disrupted by a binary-single dynamical interaction (DBS) or binary-binary interaction (DBB). 
EXBS, EXBB, DBS and DBB cannot be initial types for BSS. Initial BSS type can be EM, EMT, ED, CSS, CBS or CBB, and only later BSS can change its type into another one.

Similar distinction for BSS channels of formation and type changes are introduced also by \citet{Leonard1996ApJ...470..521L}. Abbreviations of ours channels of formation and type changes one can find summarized in Tab.~\ref{tab:Abbreviations}.

\begin{table}
	\begin{center}
	    \begin{tabular}{ | p{3cm} | p{4cm} }
	    Abbreviation  & Description \\
	    \hline
	    
EM & Evolutionary Merger\\
EMT & Evolutionary Mass Transfer\\
ED & Evolutionary Dissolution\\
CSS & Collision Single-Single\\
CBS & Collision Binary-Single\\
CBB & Collision Binary-Binary\\
EXBS & Exchange Binary-Single\\
EXBB & Exchange Binary-Binary\\
DBS & Dissolution Binary-Single\\
DBB & Dissolution Binary-Binary\\
	    \end{tabular}
	\end{center}
	\caption{Abbreviations of channels of formation of BSS and type changes}
	\label{tab:Abbreviations}
\end{table}

\subsubsection{Evolutionary mass transfer BSS}
\label{sec:EvolutionMassTransferBSS}

For BSS from a one channel of formation there could be different physical processes which are responsible for creating such a BSS. Thus, in this subsection BSS from the channel EMT will be split further into several subgroups and their properties will be discussed.

\begin{figure*}
	\centering  
		\includegraphics[width=59mm,angle=270]{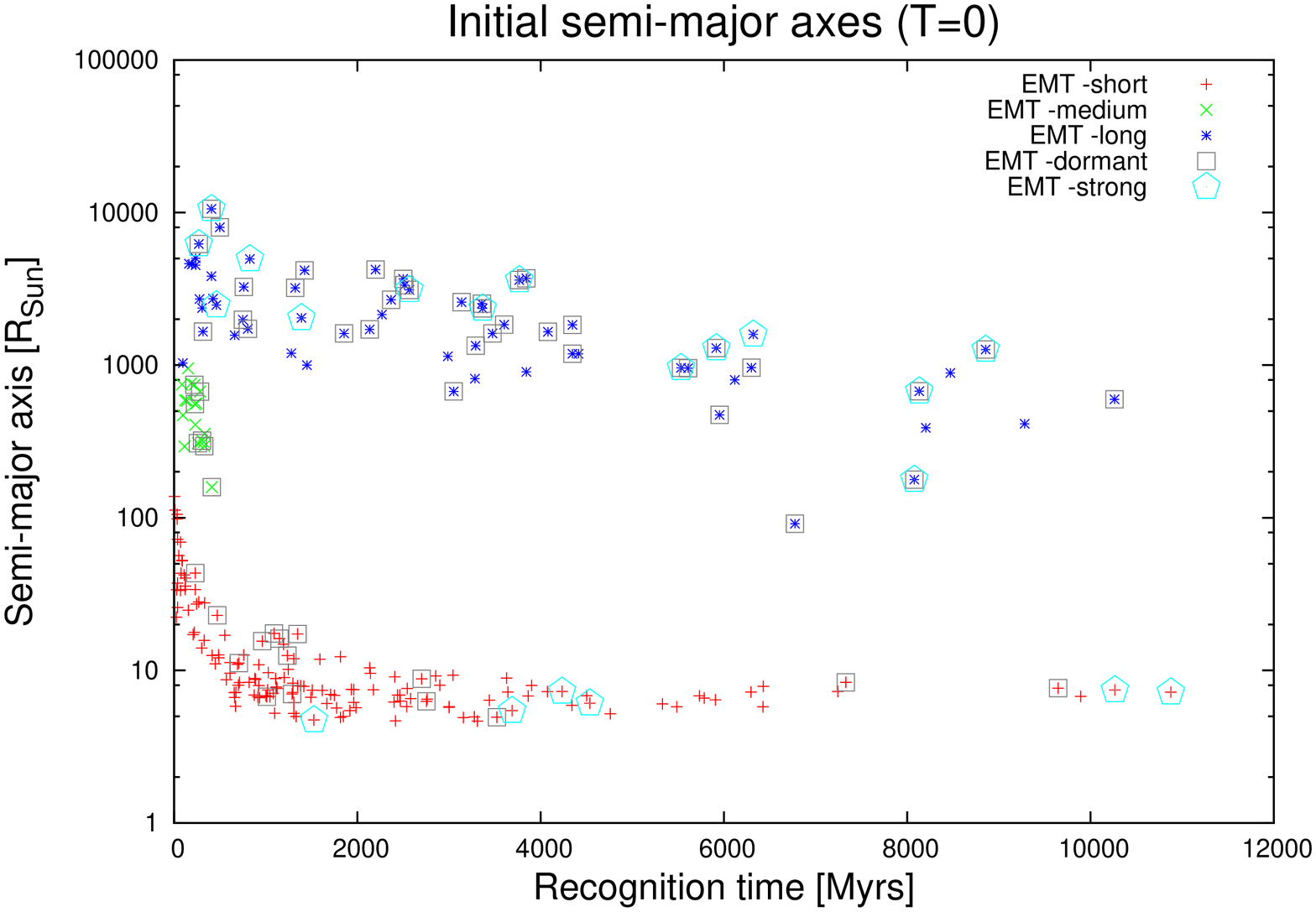}  
		\includegraphics[width=59mm,angle=270]{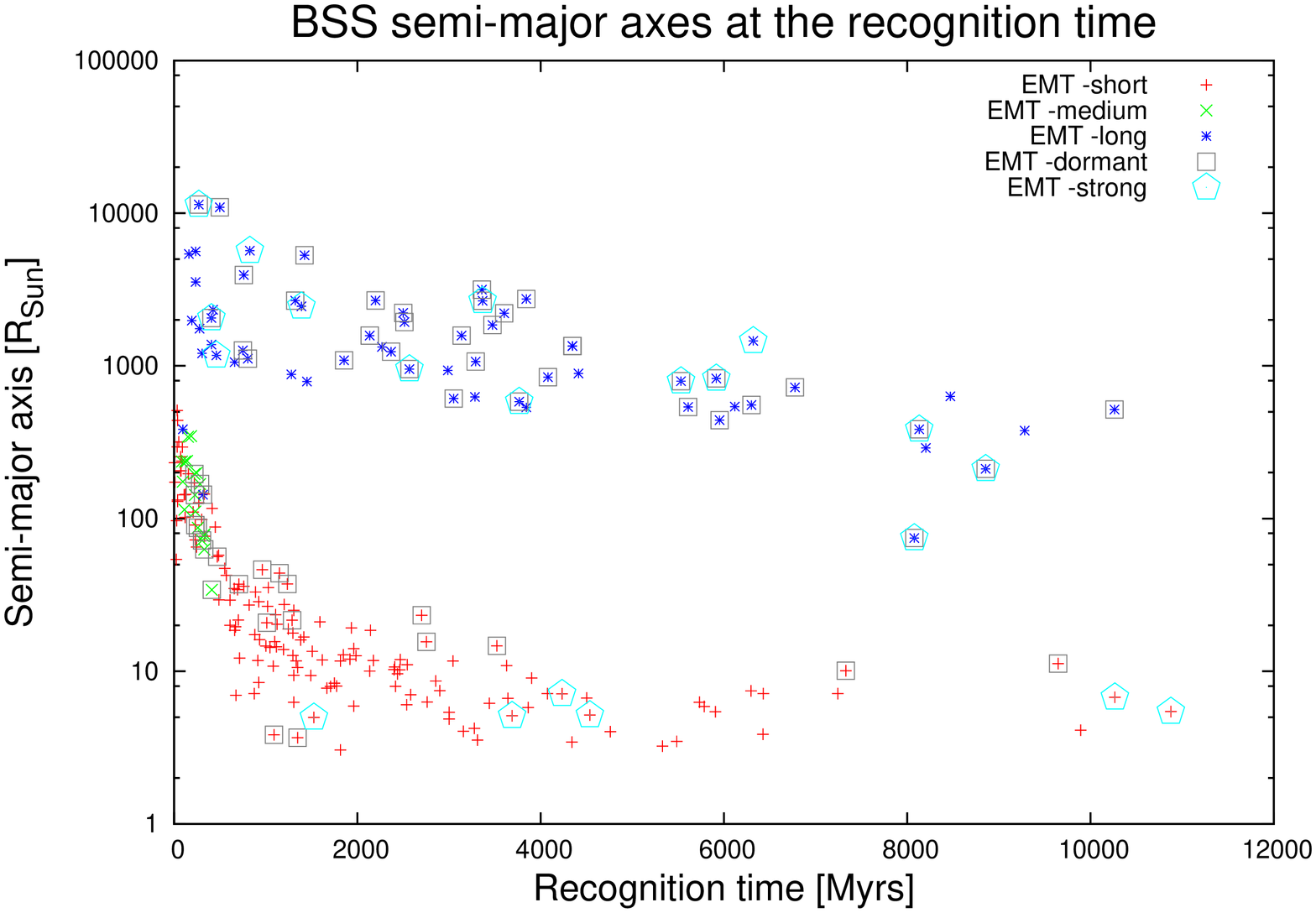}
		\includegraphics[width=59mm,angle=270]{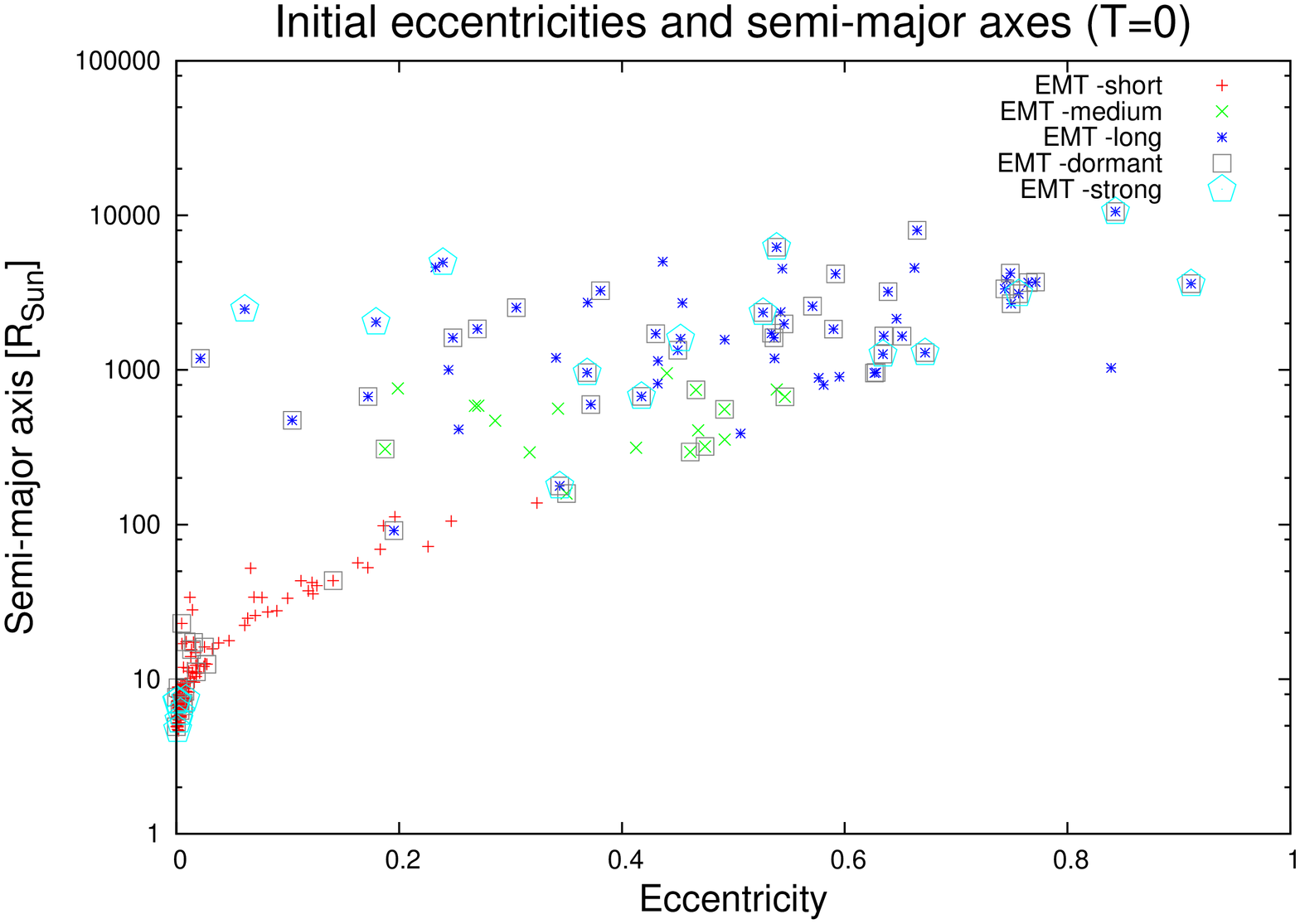}
		\includegraphics[width=59mm,angle=270]{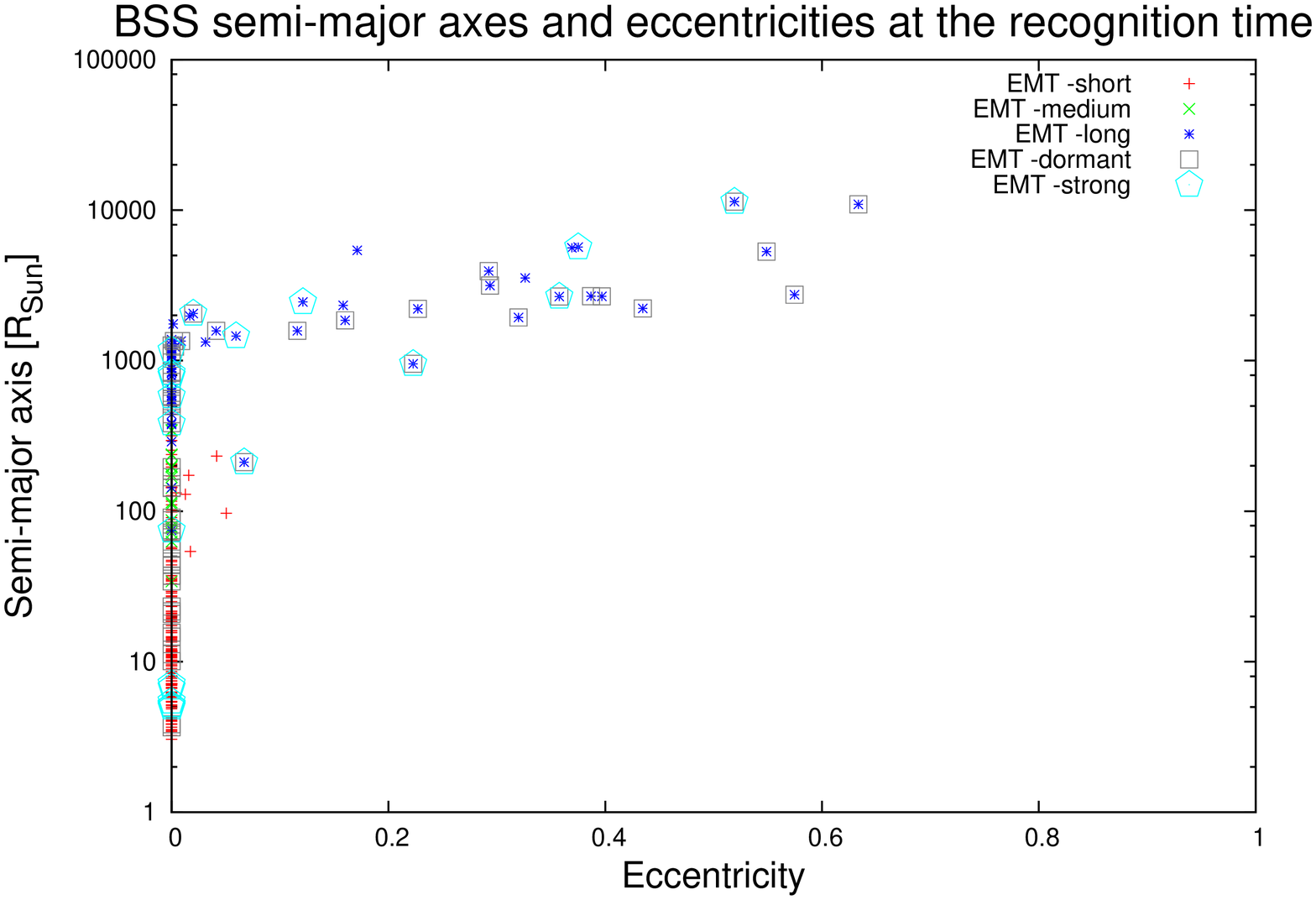}
  		\includegraphics[width=59mm,angle=270]{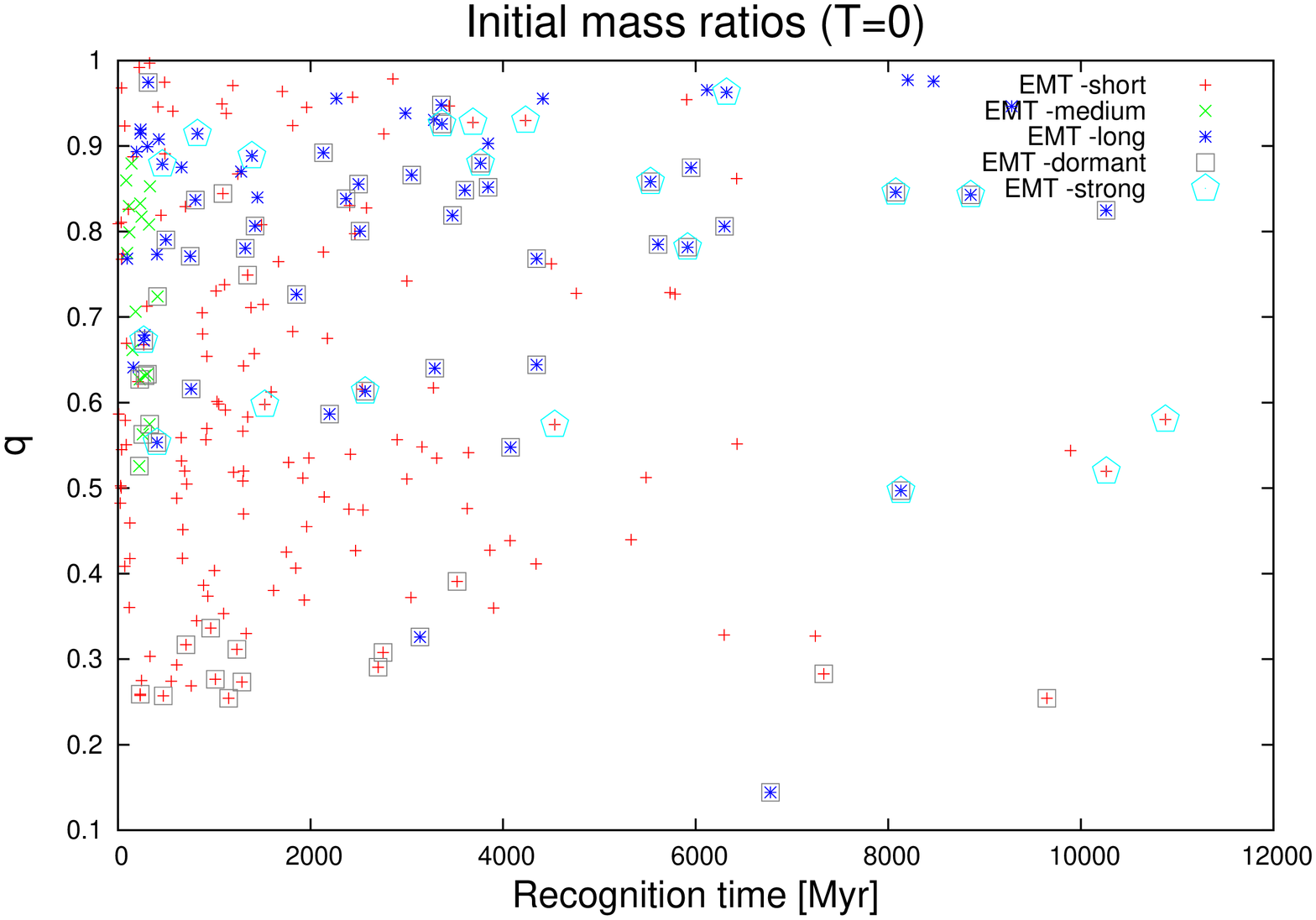}
		\includegraphics[width=59mm,angle=270]{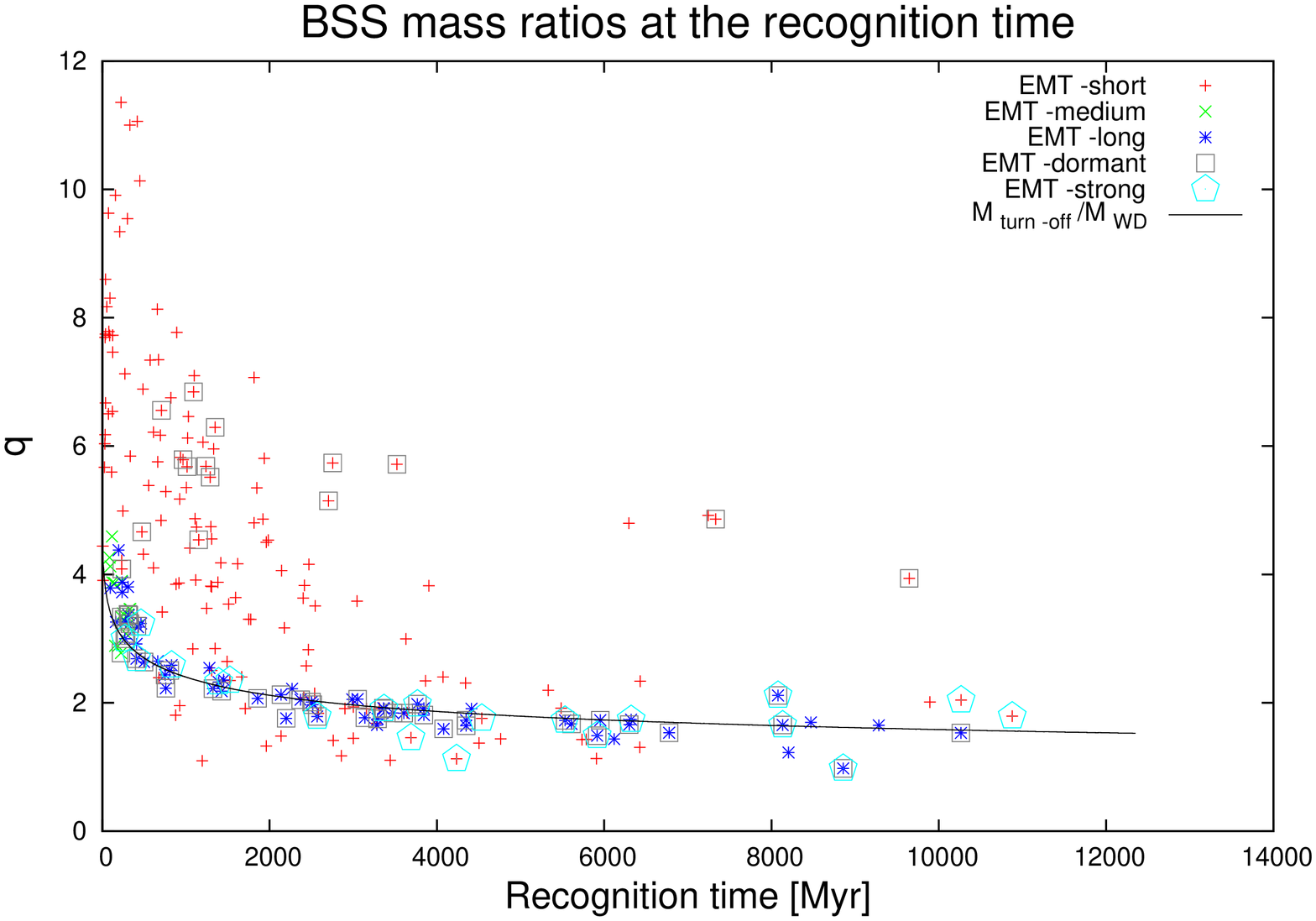}
		\caption{Properties of BSS from the channel EMT divided into 3 separate subgroups: short (red), medium (green) and long (blue) period EMT. Additionally, there are introduced here two subgroups: one with dormant EMT and one with EMT which had at least one strong dynamical interaction before becoming BSS (see text for their definitions). 
		The top-left panel shows semi-major axes [$\mathrm{R_{\odot}}$] of initial binaries, in which EMT were created later on, with positions on the X axis set according to BSS recognition time. The top-right panel shows semi-major axes [$\mathrm{R_{\odot}}$] of EMT in the time of their recognition. 
		In the middle panels there are semi-major axes and eccentricities similarly at the time T=0 and at the time of BSS recognition. 
		On the two bottom panels there are the mass ratios at the time T=0 (bottom-left), and at BSS recognition time (bottom-right). 
		On the bottom-right panel there is also one fit curve ($M_{turn-off} / M_{WD}$), described in details in the text. 
		In general, left panels show initial properties of binaries (T=0), where later on BSS were created, and right panels show properties of BSS at the recognition time.
}  
		\label{pic:EMTSubgroups}
\end{figure*}

Fig.~\ref{pic:EMTSubgroups} shows properties of BSS from the channel EMT divided into 3 separate subgroups: short, medium and long period EMT. Each subgroup corresponds to a different initial semi-major axes range (see red, green and blue points on the top panels in Fig.~\ref{pic:EMTSubgroups}). Additionally, there are introduced two subgroups: dormant and strong BSS. They actually consist of points from first 3 subgroups. Dormant BSS are BSS for which there is some delay between the last mass transfer and a time when star actually exceeded the turn-off mass (for details about lifetimes of BSS see Sec.~\ref{sec:LifetimesOfBlueStragglers}). The time when the last mass transfer occurred is called the creation time of a BSS, because such an event is the real cause responsible for the creation of a BSS. The time when a BSS actually exceeded the turn-off mass is called the recognition time. Furthermore, the time range between the creation time and the termination time of a BSS is called the total lifetime of a BSS. The time range between the recognition time and the termination time of a BSS is called the effective lifetime of a BSS. Strong BSS are those which had at least one strong dynamical interaction before BSS creation time (binary's properties had to be changed at least by 10\%, see later Sec.~\ref{sec:InducedMT}).

In the top two panels in Fig.~\ref{pic:EMTSubgroups} one can see how semi-major axes change from the initial one (at the time $T = 0$) to the semi-major axes at the recognition time. At the recognition time semi-major axes create two separate trends (the top-right panel in Fig.~\ref{pic:EMTSubgroups}). The long period EMT belong to the first one and short and medium period EMT to the second one. After a detailed analysis of the history of formation of these groups of EMT it came up that there are two different scenarios responsible for their creation. 

\subsubsection*{Long period EMT}

A donor star at some point starts to leave the main sequence, enters the giant branch and ejects some mass through stellar winds. A star, which later on becomes a BSS, gains mass of about a few $\sim 0.01 \mathrm{M_{\odot}}$, which is not enough to become a BSS. The donor star goes further through the AGB phase, which means that it ejects outer layers at a rate of about $10^{-6} - 10^{-7}$ $\mathrm{M_{\odot}}$ per year. This stage can last at most few Myrs before the donor star becomes a white dwarf surrounded by a planetary nebula. The future BSS increases its mass by about $< 0.1 \mathrm{M_{\odot}}$ through stellar winds and by moving through the ejected envelope. Initial eccentricities are typically larger than 0.1 (up to 0.9) which makes the mass transfer a bit easier. Although a circularization occurs, it does not change eccentricities to 0.0 for all binaries. Some of them at the recognition time are still larger than 0.0. The donor star becomes a white dwarf, but the star which gained the mass in many cases still does not exceed the turn-off mass. Even if the mass of the star, which gained mass, is close to the turn-off point, it takes some time for the BSS to actually exceed the turn-off mass. This is why long period EMT are in majority dormant BSS. It takes sometimes even several Gyrs until the star finally exceeds the turn-off mass and becomes BSS. What is also important, is that almost all of long period EMT are short living BSS (see Sec.~\ref{sec:LifetimesOfBlueStragglers}). It means that BSS gained only a small amount of additional mass from an evolving companion and exceeded the turn-off mass, as a MS star, just for the short period of time before it left the MS (at most a few hundred Myrs for time $T > 2$~Gyrs). Lifetimes of BSS are described in details in the Sec.~\ref{sec:LifetimesOfBlueStragglers}.

Semi-major axes of binaries with long period EMT are getting smaller with the recognition time (see the top-right panel in Fig.~\ref{pic:EMTSubgroups}). It is caused by the fact that the mass transfers are getting smaller with time too. In order to transfer enough mass to create a BSS, binaries need to have smaller semi-major axes.

Many of the long period EMT are also BSS which had some strong dynamical interactions before becoming a BSS. It is most likely because of the fact that they have large semi-major axes and it is simply easier to change their properties by at least 10\% in comparison to the short period EMT.

\subsubsection*{Short and medium period EMT}

Short period EMT are created by a mass transfer in a form of a Roche lobe overflow. When a donor star starts to evolve from the main sequence its radius increases. At some point the binary becomes semi-detached and a conservative mass transfer begins. If the more massive star loses mass, then the semi-major axis of the binary is getting smaller. The mass transfer lasts typically for several hundreds of Myrs until the mass ratio flips in favor of the future BSS. For some cases, when the mass transfer continues also when the more massive star becomes less massive, the semi-major axis increases. Thus, in the top-right panel in Fig.~\ref{pic:EMTSubgroups} the semi-major axes are more scattered than initial ones. The donor star goes through the Hertzsprung gap phase, then the giant branch phase. The mass transfer stops and the donor star ends up as a helium core star. In many cases after some time such a binary merges into one star, which is not a BSS any longer. This mechanism is possible only for compact binaries (see the top-right panel in Fig.~\ref{pic:EMTSubgroups}).

Semi-major axes of short period EMT at the recognition time decrease with time (from several hundreds $\mathrm{R_{\odot}}$ to about 10 $\mathrm{R_{\odot}}$ after about 1~Gyrs). It is caused by the fact that for less heavier MS stars also radii are smaller. Thus, when one of the stars of a binary leaves the main sequence, the other one has to be sufficiently close to create the Roche lobe. Additionally, as cluster evolves, there are less and less binaries which have parameters needed to create the Roche lobe overflow. 

On the top-right panel one can see that medium period EMT (green points) shrink their semi-major axes so that they overlap with short period EMT. In turn, short period EMT (red points) for time $< 500$~Myrs increase their semi-major axes. These are the only subgroups of EMT which change their semi-major axes that much from the beginning of the evolution. It is caused by the fact that for the medium period EMT, during a heavy mass loss through stellar winds, some mass is ejected outside the binary so that the overall binary mass is smaller. A decrease of the mass of the binary causes a shrinking of the semi-major axes. 
Eventually, a semi-major axis decreases to such a value that the short semi-detached phase in the binary starts. Mass is transferred via the Roche lobe overflow to the future BSS (this time more mass than the mass gained from previous stellar winds) and the star becomes a BSS. In turn for short period EMT, when a mass transfer occurs via the Roche lobe overflow in the semi-detached phase, the overall mass of a binary is conserved. At some point, the star which was loosing mass, becomes less massive (Algol-type star). The mass ratio inverts, but the donor star which is less massive, still looses its mass. Because of the conservation of the momentum, the semi-major axis increases. As a result, medium and short period EMT overlap each other at the recognition time.

The gap between short and long period EMT is not caused by the lack of binaries. There are still binaries for the whole range of semi-major axes, and there are still mass transfers through mechanisms described above. However, either stellar winds are not large enough to exceed the turn-off mass, or stars in a binary are too far of each other to create the Roche lobes overflows. In these cases the mass transfers are not sufficient enough to create EMT BSS. For some other cases the mass transfers are significant ($>~0.1 \mathrm{M_{\odot}}$) but the overall mass of the star does not exceed the turn-off mass. The turn-off point during the whole simulation does not decrease enough to allow these stars to become a BSS.

\subsubsection*{Circularization of the orbits}

On the middle panels in Fig.~\ref{pic:EMTSubgroups} there are eccentricities and semi-major axes for all subgroups of EMT. Almost all eccentricities for short and medium period EMT were changed due to binary stellar evolution to $\cong~0.0$ (circularization). 
Circularization for them is caused by tidal forces \citep{Hurley2002MNRAS.329..897H}. 
Only a few of them are not entirely circularized. These short period EMT which at the recognition time have eccentricities $> 0.0$ are created during the first few dozens Myrs, so there is simply not enough time to circularize their orbits before BSS phase.

Eccentricities for long period EMT became smaller, some of them are changed to 0.0. Circularization for long period EMT is caused by stellar winds \citep[Eq. 15]{Hurley2002MNRAS.329..897H}. For the widest binaries ($> 10^3 \mathrm{R_{\odot}}$) orbits have still high eccentricities. They decrease in general, but not entirely.

\subsubsection*{Masses}

Mass ratios for subgroups of EMT are shown in the two bottom panels in Fig.~\ref{pic:EMTSubgroups}. These mass ratios are for all cases calculated as mass of a BSS over a mass of a companion star (the bottom-right panel). On the bottom-left panel there are initial mass ratios for binaries, which later on created EMT. 

Initial mass ratios are mostly spread for short period subgroup of EMT (from 0.25 up to 1.0). Short period EMT have the mass transfers through the Roche lobe overflows. 
When stars are close to each other, the mass transfer through the Roche lobes is more efficient. The companion looses a lot of mass, mass transfer lasts longer, and the companion cannot create a degenerated core. Thus, the mass ratios for such BSS are more spread. In the majority of cases the companion becomes a helium core star. Only in several other cases the companion evolves to WD. Thus, many of short period EMT at the recognition time have high mass ratios (up to $\sim 12$). 

Initial mass ratios (at the time T=0) for medium and long period EMT occupy similar regions: from $\sim 0.5$ up to 1.0. 
The mass ratios at the recognition time drops from about 4 (at the time T=0) to about 1.5 (at the time 10~Gyrs). They have a narrow range at the recognition time because the companions evolve to WDs in almost all cases. Long period EMT can gain only small amount of matter from a companion (typically 0.1 $\mathrm{M_{\odot}}$ after 500~Myrs). Thus, for long period EMT mass ratios at the recognition time create a distinct trend (see the bottom-right panel in Fig.~\ref{pic:EMTSubgroups}). We plot there a black curve based on the following equation:

\begin{equation}
	q = M_{turn-off} / M_{WD}
	\label{eq:LongEMTQ}
\end{equation}
where $M_{turn-off}$ is the turn-off mass and $M_{WD}$ is a mass of a white dwarf \citep[Tab. 1]{Chernoff1990ApJ...351..121C}.
This equation shows how mass ratio changes through the entire simulation for long period EMT. The masses of BSS are just slightly larger than turn-off mass, thus in the nominator there is $M_{turn-off}$. WDs are companions in the long period EMT, thus in the denominator there are masses of WDs calculated based on \citet[Tab. 1]{Chernoff1990ApJ...351..121C}. One can see that the black curve in the bottom-right panel in Fig.~\ref{pic:EMTSubgroups} reproduces long period EMT mass ratios very well. They have a narrow range and predictable values through the entire simulation.

\subsubsection{Evolutionary merger BSS}
\label{sec:EvolutionMergersBSS}

\begin{figure}
	\includegraphics[width=\picWidth,angle=270]{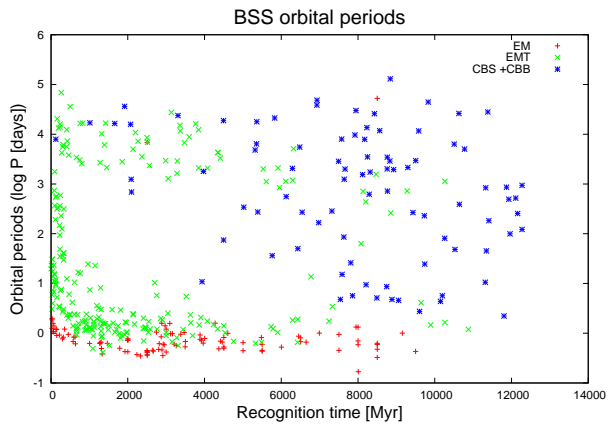}
	\caption{BSS orbital periods at the recognition time for different channels of formation}
	\label{pic:CreationTime_vs_BSSOribtalPeriod}
\end{figure}

Fig.~\ref{pic:CreationTime_vs_BSSOribtalPeriod} shows last orbital periods before a star was recognized as a BSS as a function of the recognition time. EM BSS are compact objects, with orbital periods $\sim~1$ day or less and circular, or almost circular, orbits. They were created also from the compact binaries with semi-major axes, at the $T=0$, from about $10^{0.5}$ up to about $10^{1.2} \mathrm{R_{\odot}}$ (see bluest points in the density plot in Fig.~\ref{fig:HistInitialConditions}) and with eccentricities $\sim~0.0$. These are expected properties for EM BSS, because in order to have an evolutionary merger, there has to be a very close binary. 

For the whole simulation there were observed only two exceptions for EM channel with the orbital periods $\sim~10^4$ days. After checking the history of these long period EM it came up that these were binaries which were changed by interactions, got very high eccentricities ($>~0.99$) and after some time the binary merged into one star creating BSS. 

\begin{figure}
	\includegraphics[width=\picWidth,angle=270]{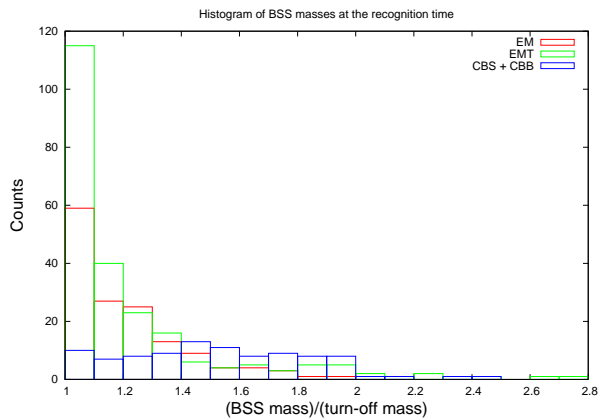} 
	\caption{Histogram with masses of BSS divided by the turn-off mass at the recognition time for different channels of formation}
	\label{pic:BSS_hist_bin_and_sin_masses2}
\end{figure}

Fig.~\ref{pic:BSS_hist_bin_and_sin_masses2} shows masses of BSS at the recognition time for 3 major channels of formation: EMT, EM and CBS+CBB. The majority of EM BSS have masses which are just slightly above the turn-off mass. Almost all of such EM ($1 - 1.1$~$M_{turn-off}$) are the dormant EM (see Sec.~\ref{sec:LifetimesOfBlueStragglers}). When dormant EM are detected, they have masses just slightly larger than the turn-off mass. For larger EM masses ($> 1.1$~$M_{turn-off}$) there are less and less EM. 
In order to create an EM with mass close to 2~$M_{turn-off}$ both components of a binary would have to have masses close to the $M_{turn-off}$. There is simply less compact binaries with initial mass ratios close to 1. For compact binaries IMF generates rather different masses of components. However, it is still possible to have EM BSS with higher mass ratios, which eventually create heavier BSS (up to 2 times the turn-off mass). 
 
In Fig.~\ref{pic:CreationTime_vs_BSSOribtalPeriod} one can see that EM BSS consist actually of two subgroups. The first group represents EM with the recognition times $\leq 3$~Gyrs for which orbital periods clearly decrease with time. At the beginning of the simulation orbital periods of binaries, which later on become BSS, are slightly larger than 1 day. Just before 3~Gyrs orbital periods, which give rise to the number of BSS, shorten to about $10^{-0.5}$ days. Around 3~Gyrs they increase and stay around 1~day for the rest of the star cluster evolution. We carefully checked the history of the creation for many BSS of these two groups and we were able to find two different physical scenarios responsible for the creation of BSS. 

In the first group of EM ($\leq 3$~Gyrs) the typical physical process responsible for the creation of EM BSS is a merger in a binary due to the Roche lobe overflow. Typical formation scenario starts with the binary with masses like e.g. $m_1 = 3.3 \mathrm{M_{\odot}}$, $m_2 = 0.5 \mathrm{M_{\odot}}$, the semi-major axis $a = 6.2 \mathrm{R_{\odot}}$, and the eccentricity $e = 0.004$. The semi-major axis decreases with time but only slightly, the eccentricity circularizes to 0.0, and after a few hundreds Myrs, when the heavier star increases its radius, a semi-detached phase starts. It lasts less than 1~Myr and finally the binary merges into one BSS star. 

The typical scenario for EM BSS created after 3~Gyrs involves the magnetic braking which eventually leads to a merger event in a binary. The magnetic braking affects stars with masses less than $1.25 \mathrm{M_{\odot}}$ (for details see \citet{Hurley2002MNRAS.329..897H}). Indeed, at the time $T~\sim~3$~Gyrs the turn-off mass equals $1.25 \mathrm{M_{\odot}}$ and the magnetic braking starts to work for both components in binaries with MS stars. Typical formation scenario for EM BSS recognized after 3~Gyrs is as follows. The initial binary properties are e.g. $m_1 = 0.88 \mathrm{M_{\odot}}$, $m_2 = 0.13 \mathrm{M_{\odot}}$, $a = 8.89 \mathrm{R_{\odot}}$, $e = 0.003$. The masses and the semi-major axis are in general smaller in comparison to the binaries which created EM through the Roche lobe overflow. The semi-major axis decreases after 8~Gyrs up to $a = 3.41 \mathrm{R_{\odot}}$, the eccentricity circularizes to 0.0. The stars merge into one object, without detecting the semi-detached phase by the MOCCA code. This and larger decrease of the semi-major axis (through the magnetic braking) are two differences between these two groups of EM BSS. 

Additionally, these two groups of EM BSS still overlap because magnetic braking works for binaries with stars' masses below $1.25 \mathrm{M_{\odot}}$, and there are such binaries in the system before 3~Gyrs. It is more complicated also because of the existence of a dormant EM BSS for which the merger event can be significantly earlier than the recognition time (see Sec.~\ref{sec:LifetimesOfBlueStragglers}). 
EM BSS created by the Roche lobe overflow are created mainly for the more massive stars in the first few Gyrs and those created by the magnetic braking are created mainly for the later stages of the star cluster evolution. 
But still, there are some EM created by magnetic braking before 3~Gyrs (for masses less than $1.25 \mathrm{M_{\odot}}$), and there are some Roche lobe overflow EM BSS created after 3~Gyrs. It will be interesting to see if a similar division is present also for different initial binary properties. We plan to check this in details in the next paper.

\subsubsection{Collisional BSS}
\label{sec:DynamicalMergersBSS}

Collisional BSS (CBS and CBB) are created in strong dynamical interactions between binaries and single stars. Fig.~\ref{pic:CreationTime_vs_BSSOribtalPeriod} shows orbital periods of binaries at the recognition time. Before core collapse ($< 6$~Gyrs) CBS and CBB were created from binaries with rather large orbital periods. It is caused by the fact that before the core collapse, when density is smaller, only very wide binaries have probabilities large enough to have dynamical interactions. Additionally, some of such primordial binaries are already in the core due to the initial conditions. During and after the core collapse, when the core is getting denser, there are more CBS and CBB in general. They are created also from more compact binaries with orbital periods of about a few up to $10^5$ days. After the core collapse, the density in the core is high enough to create BSS, even from close binaries. We checked also the initial properties of the binaries, which later on created CBS and CBB, and found no correlations with the orbital periods. Properties of such primordial binaries are uniformly distributed for the whole range of the semi-major axes and eccentricities in the histogram in Fig.~\ref{fig:HistInitialConditions}.

Fig.~\ref{pic:BSS_hist_bin_and_sin_masses2} shows masses of BSS divided by the turn-off mass. CBS and CBB have roughly uniformly distributed masses from 1 up to 2 times the turn-off mass. There is no distinct peak in the distribution of the masses for this channel. It means that CBS and CBB BSS are formed from various MS stars and there is no particular preference which of them collide in dynamical interactions. The only limitation is that two MS stars can form BSS with the mass equal to at most twice the turn-off mass. However, there are still several CBS or CBB with masses exceeding the turn-off mass more than two times. These BSS were created by rather complex scenarios. In some cases there were more than one merger for a star which became a BSS. In other scenarios, there was an additional mass transfer from the companion to the already existing BSS, which gave BSS enough mass to exceed the turn-off mass more than two times. 

Initial binaries' mass ratios for CBS and CBB channels are uniformly distributed up to 2.5 for various eccentricities, up to 0.8. CBS and CBB become important after several Gyrs, and during that time there are actually no more binaries with mass ratios larger than about 3. CBS and CBB binaries' mass ratios more or less cover overall mass ratios of all binaries in the system at a given time. There is no visible preference which binaries could create CBS or CBB in collisions during dynamical interactions.

Apart from to the fact that CBS and CBB were created in dynamical interactions, there are no distinct physical processes or typical formation scenarios for creating these types of BSS. Collisions occurred for binaries with various semi-major axes and eccentricities. Some dynamical interactions were simple, other ones were resonant.
For some other cases, during the dynamical interactions, the collision occurred for a binary after it already passed the other object (a binary or a single star) at the pericenter. The binary, going through the pericenter, gained a very high eccentricity and a collision in a binary occurred while it was already moving away from the other object. But still such dynamical interactions were not resonant because the scattering objects run through the pericenter only once.

\subsubsection{Types changes}
\label{sec:TypeChanges}

\begin{figure}
\centering 
\begin{minipage}{.4\textwidth}
	\includegraphics[width=\picWidth,angle=270]{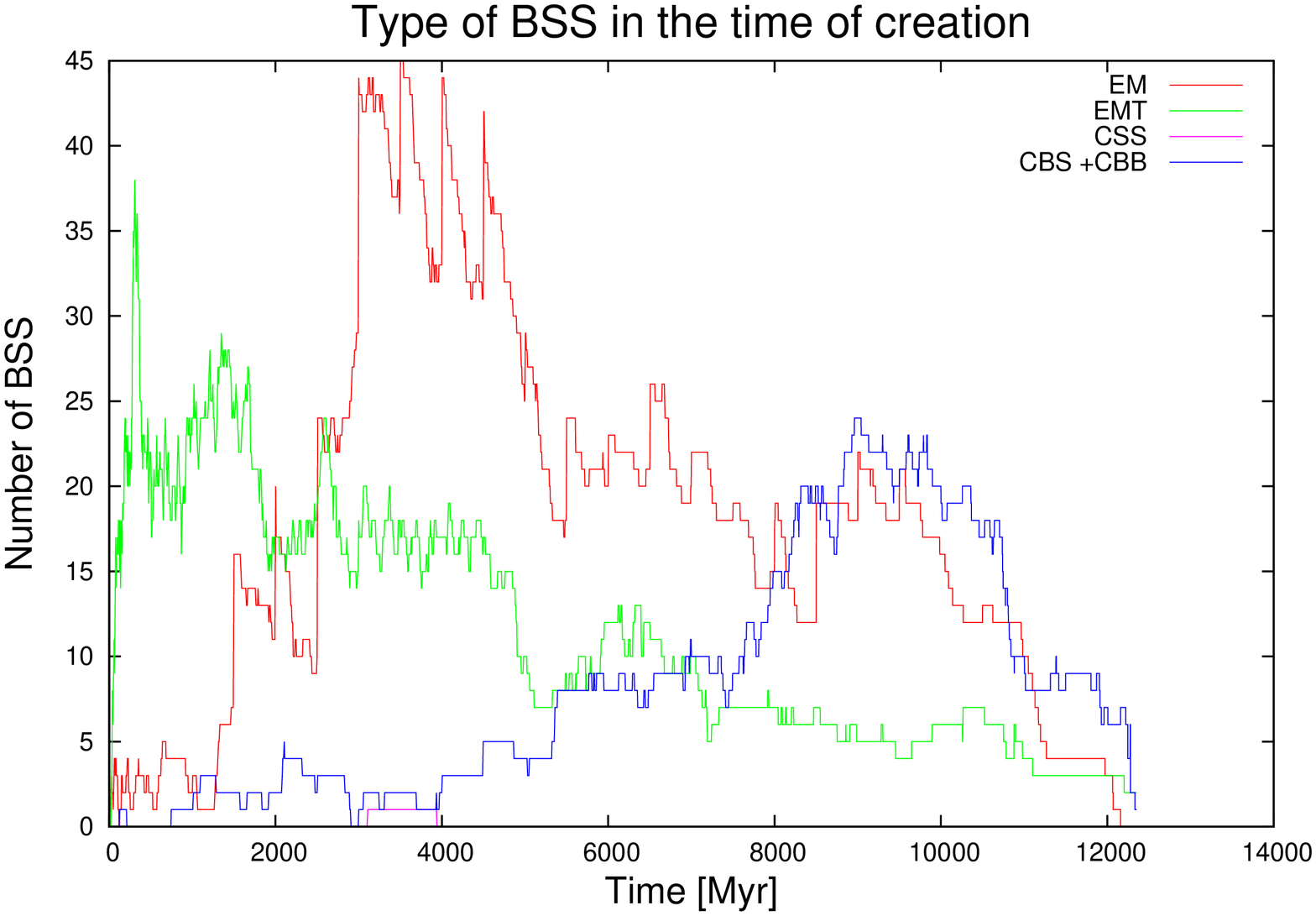} 
	\caption{BSS types at the creation time as a function of the time}
	\label{pic:Time_vs_BSSInitialType}
\end{minipage}
\begin{minipage}{.4\textwidth}
	\includegraphics[width=\picWidth,angle=270]{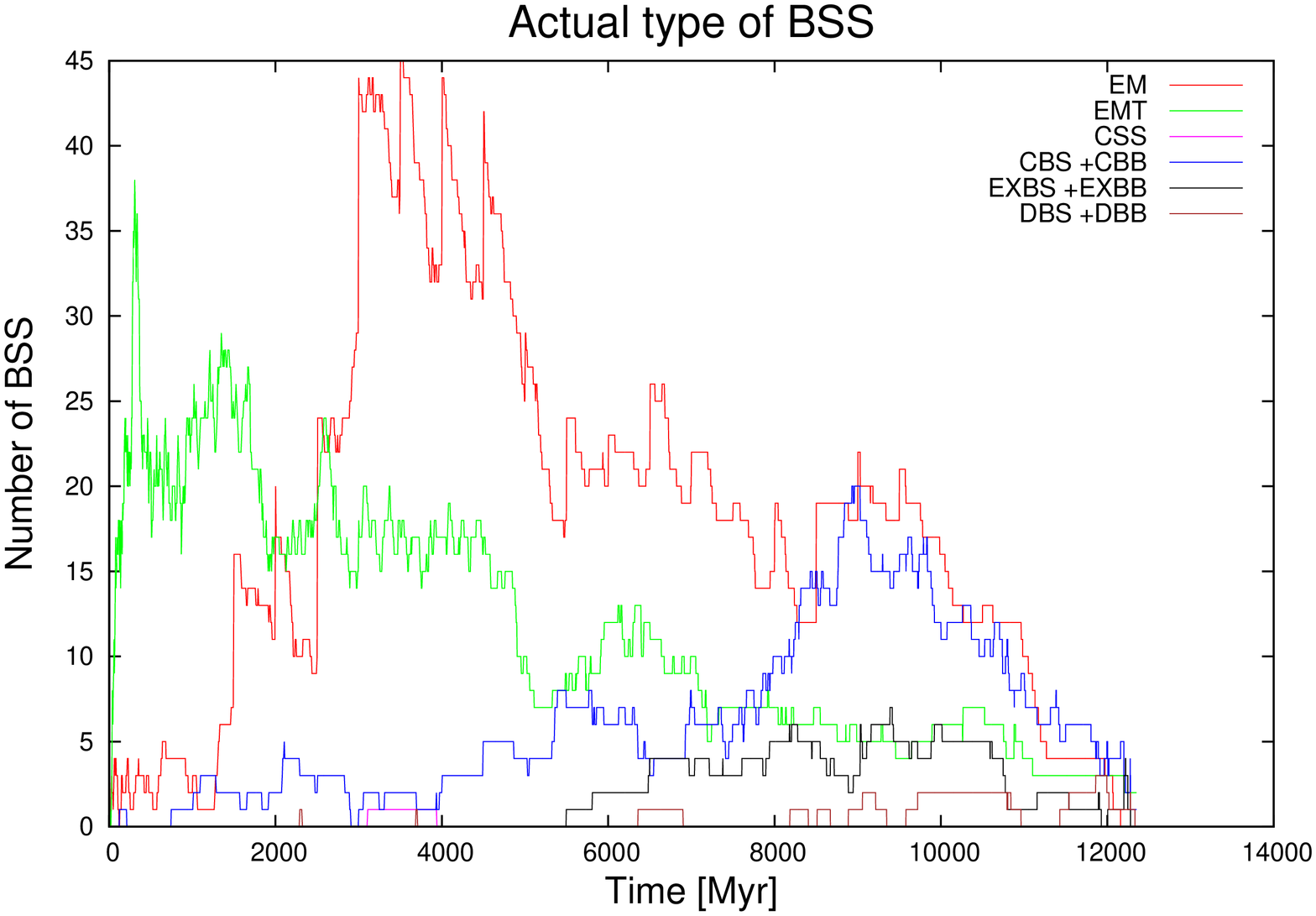}
	\caption{Actual types of BSS as a function of the time}
	\label{pic:Time_vs_BSSActualType}
\end{minipage}
\end{figure}

Fig.~\ref{pic:Time_vs_BSSInitialType} shows the number of BSS as a function of time. One can see how different channels of formation of BSS change their significance during the evolution of the star cluster. EMT channel is the most active in the beginning of the simulation. In the beginning, there are simply more binaries which are able to start a mass transfer. First EMT BSS are formed in very massive binaries, which means that these binaries evolve fast from MS, there occur violent mass transfers which create short living BSS (see Sec.~\ref{sec:LifetimesOfBlueStragglers}). Stellar evolution for massive stars is very fast and after several hundreds of Myrs, when EMT gains its peak, the EMT channel becomes less and less significant as the cluster evolves. The number of BSS created by the EMT channel continuously decreases but stays active for the whole simulation. The number of EMT BSS decreases because it depends mainly on the initial properties of the binaries. Depending on the initial mass, a star leaves MS at a different time. Thus, different binaries starts the EMT BSS phase at different times. As the evolution of the star cluster continues, the number of EMT BSS continuously drops, because there are less and less binaries which can start mass transfers and create EMT BSS. 

The number of EM BSS in Fig.~\ref{pic:Time_vs_BSSInitialType} starts to increase after 1~Gyrs, when a violent evolution of most massive stars is over. A binary, depending on the physical process responsible for the creation of BSS (the Roche lobe overflow or magnetic braking), needs some time to actually merge. Thus, the first increase of EM BSS appears after around 2~Gyrs. The peak of the number of EM falls to around 4~Gyrs and after that it decreases. The peak of the number of EM BSS is most likely caused by the fact that both scenarios for EM creation, the Roche lobe overflow and magnetic braking, work at that time most efficiently. 
The magnetic braking (see Sec.~\ref{sec:EvolutionMergersBSS}) works for stars with masses smaller than $1.25 \mathrm{M_{\odot}}$. Thus, around the time $T = 3$ Gyrs it starts to work for both components in binaries with MS stars.
Because of that, it is possible to create mergers more efficiently and also from slightly wider binaries. Thus, there is a distinct peak in the orbital period values for EM around 3~Gyrs (see Fig.~\ref{pic:CreationTime_vs_BSSOribtalPeriod}) and an increase of the number of EM around 3-4~Gyrs (see Fig.~\ref{pic:Time_vs_BSSInitialType}). There are simply more binaries with proper initial parameters to create EM BSS through the magnetic braking. 

CBS and CBB are created in star cluster from the beginning, but significantly more of them are created during a core collapse and after that ($> 6$~Gyrs). CBS and CBB are created mainly in the core up to the half-mass radius (see Fig.~\ref{pic:BSS_creation_time_vs_last_merger_pos}), at the time of the core collapse (8 Gyrs), and afterwards, when the core is dense, they become one of the most active channels of formation of BSS.

In Fig.~\ref{pic:Time_vs_BSSInitialType} one can see that there is only one BSS, which was created by the CSS channel (about 3~Gyrs). Just to remind, the CSS channel is the only channel which creates BSS from a collision between two single stars. All the other channels involve one or more binaries. It means, that actually almost 100\% of BSS, at least for this test simulation, were created from binaries. Even if the CSS channel is not important for our test model, it does not necessarily mean that it will be not important for other simulations with different initial conditions. Perhaps for star clusters with higher concentrations there will be more CSS?

We already mentioned in the Sec.~\ref{sec:ChannelsDefinitions} that types of BSS at the creation time can change.
For example, a binary with BSS could be disrupted, or could exchange with some other star in a dynamical interaction. In such a case the actual type of BSS at a given time of the simulation can be different. Fig.~\ref{pic:Time_vs_BSSActualType} shows how types of BSS change during the simulation for different channels of formation.
Just by quick comparison between two figures Fig.~\ref{pic:Time_vs_BSSInitialType} and Fig.~\ref{pic:Time_vs_BSSActualType} one can see that some BSS changed their types during the simulation, but it concerns mainly BSS created in interactions (CBS and CBB).
In turn EMT and EM BSS did not change significantly during the simulation. The plots with BSS at the creation time look like these with actual types of BSS for these channels (see Fig.~\ref{pic:Time_vs_BSSInitialType} and Fig.~\ref{pic:Time_vs_BSSActualType}). 

\begin{figure}
	\includegraphics[width=\picWidth,angle=270]{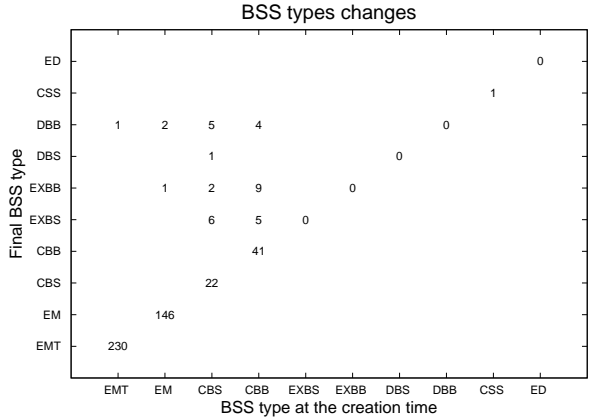} 
	\caption{BSS type changes between different BSS types. On the X axis there is the BSS type at the creation time, and on the Y axis there is the final (last known) type of the same BSS. Each number shows how many BSS changed their types from one type to another.}
	\label{pic:BSS_type_changes}
\end{figure}

Fig.~\ref{pic:BSS_type_changes} shows in details how many BSS changed their types. On the X axis there are types of BSS at the creation time and on the Y axis there is the final (last known) type of the BSS. One can see that only a few BSS which were formed as EM or EMT changed their types. There was only one EM which got into a binary (EM $\rightarrow$ EXBB). There were also two cases where EM first, as s single star, got into a binary during some dynamical interaction, and afterwards such a binary was disrupted (EM $\rightarrow$ DBB). There was only one case when EMT was disrupted (EMT $\rightarrow$ DBB).
Despite these interesting scenarios, from statistical point of view, BSS type changes for channels EM and EMT are not significant. In the next paper, we plan to study BSS population depending on many different initial conditions, both for binaries and global star cluster properties. It will be interesting to see how type changes will look like for other models. For example, for star clusters with higher concentrations, we expect to have more BSS type changes for EM and EMT channels.

BSS from EM and EMT channels do not change their types significantly during the star cluster evolution. Some of them were created in central regions, where probability of dynamical interactions with other objects is higher (see Fig.~\ref{pic:BSS_creation_time_vs_last_merger_pos}). However, even with higher dynamical interaction probabilities, there are only several EM and EMT which changed their types (see Fig.~\ref{pic:BSS_type_changes}). 
EMT do not change types probably because of the fact, that most of them are very hard binaries (see Fig.~\ref{pic:CreationTime_vs_BSSOribtalPeriod}), so there is needed another massive binary or a star and small impact parameter to disrupt or exchange BSS. Most likely, these binaries are hard enough to survive such interactions. For other EMT BSS, which have larger orbital periods (see Fig.~\ref{pic:CreationTime_vs_BSSOribtalPeriod}), the reason why they were not disrupted, is that these EMT live shortly (max $\sim 200$~Myrs, see Sec. \ref{sec:LifetimesOfBlueStragglers}), so there is simply not enough time to have strong dynamical interactions. Some of the long period EMT are found also around half-mass radius and beyond, where probabilities of the dynamical interactions are extremely small. Additionally, after the core collapse, when the dynamical interactions become important, there is simply small overall number of EMT.
EM do not change types most likely because they are single stars, and the probabilities of the dynamical interactions for them is lower than for EMT. Additionally, EM are not the most massive objects in the star cluster. Their masses are, in majority of cases, less than 1.2 of the $M_{turn-off}$ for time $> 8$ Gyrs, when the dynamical interactions are most important. Thus, the dynamical interactions occur rather for massive WDs and NS. Moreover, the number of EM is small in comparison to the number of WDs or NS at the later stages of the star cluster evolution.

The only channels which were significantly changed during simulations are CBS and CBB. One can see that BSS initially created in these channels change their types to EXBS, EXBB, DBS, DBB. It shows that the dynamical interactions (exchanges and dissolutions) starts to play a significant role for CBS and CBB in the later stages of the star cluster evolution (see Fig.~\ref{pic:Time_vs_BSSActualType}). It correlates with the core collapse event, which starts at about $6$~Gyrs, and after the core collapse, at $\sim 8$~Gyrs. Most of CBS and CBB, after the dynamical interactions which created them, are still inside binaries. For the binaries there is a higher probability that dynamical interaction will occur. Additionally, BSS in binaries are heavy objects and thus, they sink to the center of the star cluster, where dynamical interactions are even more frequent. 14 of the total 36 CBS changed their types, which gives $\sim 39\%$ efficiency. Also, 18 of the total 58 CBB changed their types, which gives $\sim 31\%$ efficiency. 

During the entire simulation there were created overall 476 BSS. The most active channels of formation were EMT and EM. There were created 231 EMT BSS (49\%) and 149 EM BSS (31\%). There were 95 BSS from the CBS and CBB channels (20\%), and there was just one BSS created by the CSS channel in our test simulation. Despite the small overall number of CBS+CBB in comparison to the evolutionary BSS (EM, EMT), they can be dominant channels of formation of BSS for clusters which are in the post collapse phase. Furthermore, for different initial conditions, we expect to have also different number of EM and EMT. These two channels are a direct consequence of binary initial conditions, thus for the simulations with less initial number of compact binaries (see Sec.~\ref{sec:InitialModel}) we expect to have much less EM and EMT BSS in general. EM particularly are formed only from primordial compact binaries with $e \sim 0.0$.

\subsubsection{Possible induced mass transfer or merger}
\label{sec:InducedMT}

\begin{figure}
	\includegraphics[width=\picWidth,angle=270]{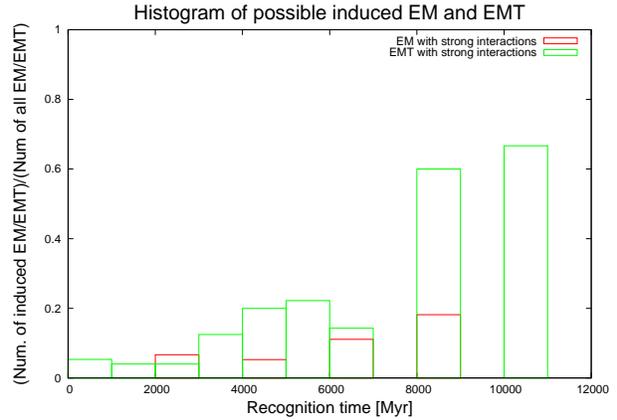}
	\caption{Histogram showing the number of BSS, binned for 1~Gyr, which might be created by induced mass transfer or induced evolutionary merger (see text for explanation). Each BSS is taken into account only once, according to its recognition time.
The value is divided by the number of all EM or EMT BSS in the particular time of the simulation, to find out whether this process is significant. Note: if there are no histogram bars for some time ranges, it means that there are no BSS with possible induced mass transfer (or merger) created for these time ranges, but there are still some BSS in the system in general.}
	\label{pic:BSSInducedEM_EMT}
\end{figure}
 
Fig.~\ref{pic:BSSInducedEM_EMT} shows EM and EMT created by a possible induced mass transfer or merger. Each histogram was normalized to 1.0 by dividing BSS number by the overall EM or EMT number of BSS in the particular time of the simulation. The possible induced EM or EMT represents BSS which had some strong dynamical interactions involving binaries, before the creation time. One can call these scenarios as possible induced mass transfers (for EMT) or possible induced mergers (for EM). Strong dynamical interactions in this context is a binary-single or a binary-binary dynamical interaction, which changed the semi-major axis or eccentricity by at least 10\%. One can see how the possible induced mass transfer could change its significance during the star cluster evolution. Induced EMT BSS in the beginning of the simulation are not significant because less than 5\% of them had some strong dynamical interactions before. Their significance increases with time as the core density increases. Later on, possible induced mass transfer increases considerably after the core collapse at 8 Gyrs, becomes important and concerns more than 50\% of all EMT BSS. BSS from the EM channel increases with time too but in this case there is less EM than EMT. EM are created from more compact binaries, so there is a smaller probability of dynamical interactions for them in comparison to EMT.

In the Sec.~\ref{sec:ComparionOldVersion} we present a simulation with the same initial conditions as for the test simulation, but with the old version of the code (without \textit{Fewbody}). The overall number of EM and EMT for both simulations, with and without \textit{Fewbody}, are essentially the same (see Fig.~\ref{pic:InitialBssNoFB}). 
It turned out that the number of possible induced EM and EMT for the simulation without \textit{Fewbody} increase with time, just like for our test simulation. The only difference is that for simulation without \textit{Fewbody} there is in general less EM and EMT with possible induced mass transfer. Next, we performed a population synthesis for the same set of primordial binaries as for our test simulation, to check how many of EM and EMT will be created (see Fig.~\ref{pic:InitialBssNoFB}). 
We found that their number is essentially the same as for the test simulation and the simulation without \textit{Fewbody}. The only significant difference in the number of EM is around 3-5~Gyrs. We were unable to determine what is the cause of this difference. The number of EM and EMT seems to not depend on the dynamical interactions. Perhaps the induced mass transfer is not important after all.  

We study in this paper only one test model. A number of possible induced mass transfer BSS appears to be important only at the later stages of the star cluster evolution -- after the core collapse. On the other hand, the numbers of EM and EMT are essentially the same for all three cases: with and without \textit{Fewbody}, and for the population synthesis (so without any dynamical interactions). It favors rather the conclusion that the dynamical interactions do not have significant influence on the creation of EM or EMT.  Perhaps possible induced mass transfer will be more significant for some specific initial conditions (star clusters with higher concentrations?). Additionally, we checked here the dynamical interactions which changed binaries' properties at least by 10\% (arbitrary chosen value). Maybe this value is too large and changes of parameters of binaries, which are smaller but frequent, should be taken into account as well? In the next paper in the series \citep{Giersz2011arXiv1112.6246G} there are compared MOCCA simulations with N-body ones. For a simulation with higher $r_{pmax}$ in the MOCCA code (higher impact parameters), there are significantly more fly-by interactions for binaries. In such a case there were also created many more BSS. This, in turn, suggests that the dynamical interactions could have some influence on the creation of BSS. For details see \citet[Fig.~7]{Giersz2011arXiv1112.6246G}. We will try to determine whether the dynamical interactions have indeed some implications for the population of EM or EMT in our future work.

\subsubsection{Comparisons with old version of the code}
\label{sec:ComparionOldVersion}

\begin{figure}
\begin{center}
  \includegraphics[width=\picWidth,angle=270]{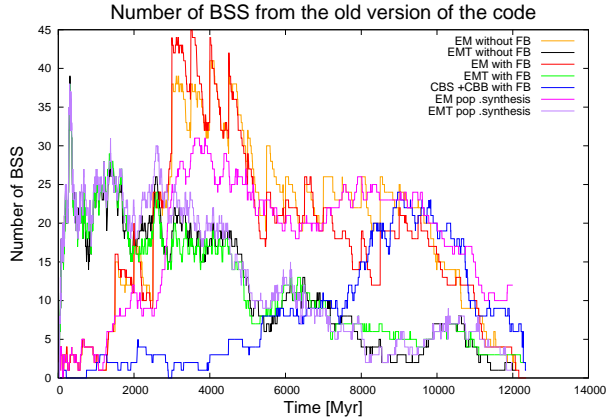}  
  \caption[]{Comparison of the numbers of BSS for the test simulation with the old version of the code (without the \textit{Fewbody}), and with the simple population synthesis (without any dynamical interactions).}
  \label{pic:InitialBssNoFB}
\end{center}
\end{figure}

Fig.~\ref{pic:InitialBssNoFB} shows the number of BSS from different channels of formation for the test simulation and for the simulation with the same initial conditions but with the old version of the code (without the \textit{Fewbody}). In the old version, instead of the \textit{Fewbody}, there are used cross-sections to determine the outcomes for the dynamical interactions. Fig.~\ref{pic:InitialBssNoFB} shows also the number of EM and EMT obtained from simple population synthesis (without any dynamical interactions).

It was mentioned before that EM and EMT BSS depend strongly on the initial binary properties. Thus, overall population of EM and EMT with and without the \textit{Fewbody} is similar (see Fig.~\ref{pic:InitialBssNoFB}). The peak value of EM is equal to about 40 BSS for both simulations, and about 25 BSS for EMT channel. In turn, for BSS created in the dynamical interactions there is a huge difference. 
The old version of the code did not create even one CBS or CBB BSS because interactions which create mergers in dynamical interactions are rather complex. The old version of the code could not deal with them in the N-body manner. 
With the \textit{Fewbody} the number of CBS and CBB outnumbers the EMT channel after 6~Gyrs. After 9~Gyrs it becomes also as efficient as EM channel. This shows how important it was to incorporate the \textit{Fewbody} into the MOCCA code in order to follow the evolution of peculiar objects, like BSS. 
For other models of star clusters, for which the dynamical interactions are even more important, one can expect even larger differences in favor of the simulations with the \textit{Fewbody}.

\subsection{BSS global parameters}
\label{sec:BssGlobalProperties}

In this section we discuss global BSS parameters, like their positions in the star cluster, lifetimes, and what is the fraction of BSS in binaries in comparison to single BSS.

\subsubsection{BSS initial and final positions}
\label{sec:BssPositions}

\begin{figure*}
	\centering
	\includegraphics[width=\picWidth,angle=270]{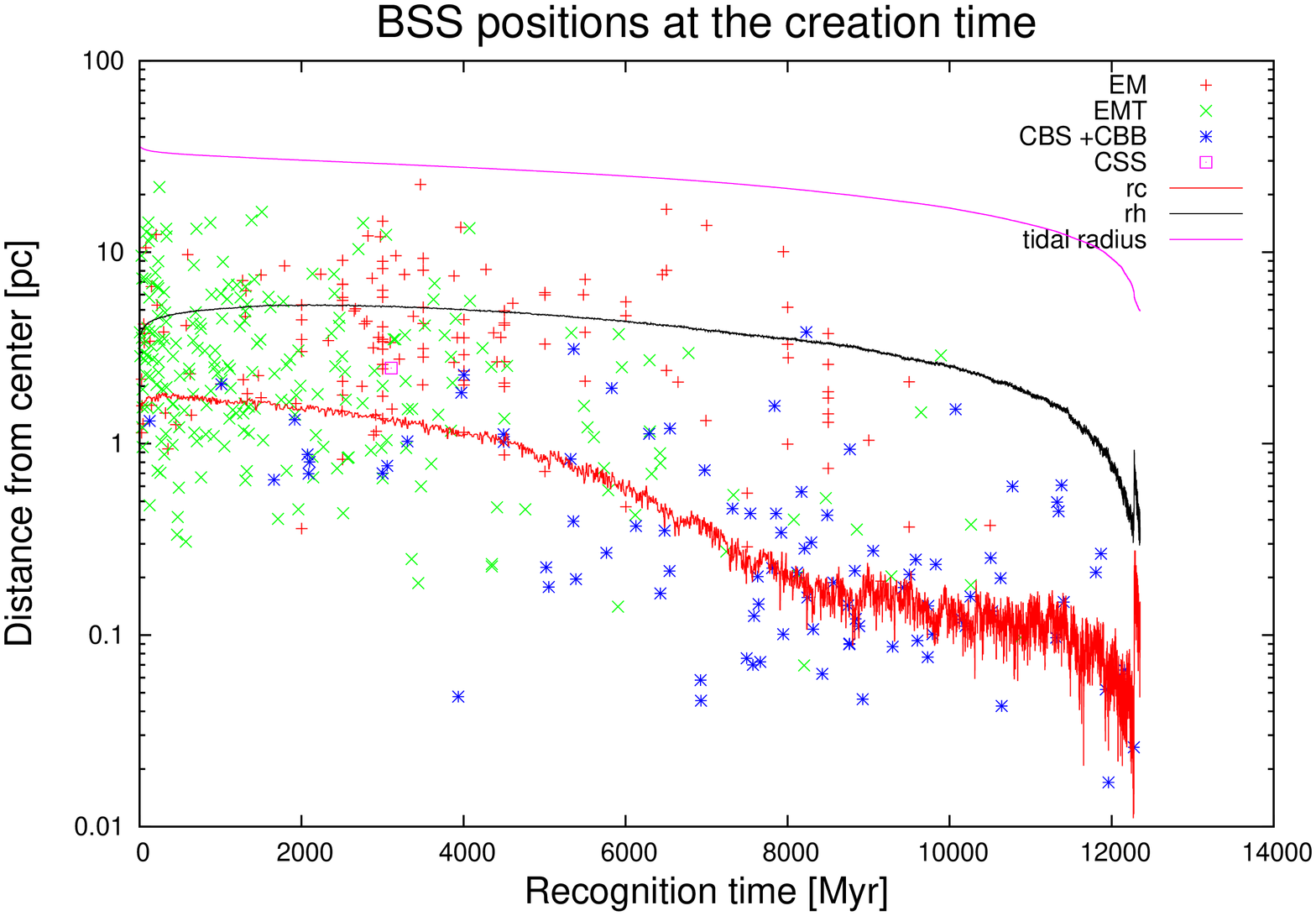} 
	\includegraphics[width=\picWidth,angle=270]{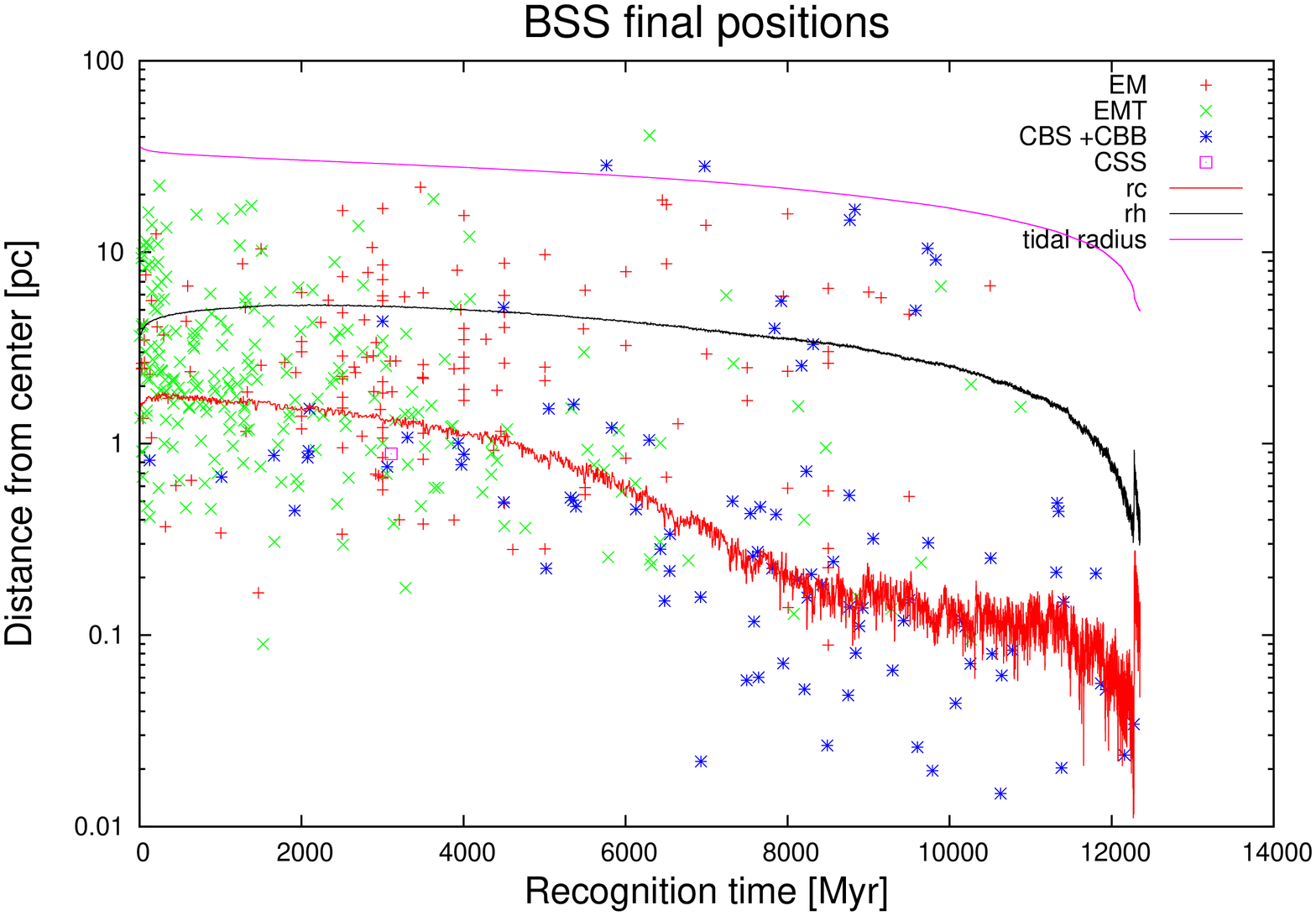}
	\caption{In the left panel there are BSS distances [pc] from the center, when last merger or last mass transfer occurred, as a function of the recognition time. In the right panel there are last known positions of BSS for different initial channels as a function of BSS recognition time. The core radius ($r_c$), the half-mass radius ($r_h$), and the tidal radius are also shown.}
	\label{pic:BSS_creation_time_vs_last_merger_pos} 
\end{figure*}       
 
Fig.~\ref{pic:BSS_creation_time_vs_last_merger_pos} shows the initial and final positions from the center of the star cluster as a function of BSS recognition time. The initial positions refers to the distances for which the last merger (for all channels except EMT), or last mass transfer (for EMT) occurred. Additionally, in Fig.~\ref{pic:BSS_creation_time_vs_last_merger_pos}, to better visualize distances, there are shown the core radius ($r_c$, calculated according to \citet{1985ApJ...298...80C}), the half-mass radius ($r_h$), and the tidal radius.

EMT BSS initial positions extends from deep inside the core up to far outside of the half-mass radius ($r_h$). Final positions for EMT are more or less the same. There is only a few EMT for which final orbits were a bit more extended. Only for one EMT the final position was located outside the tidal radius. EM BSS were created, in general, at the same distances as EMT. Final positions of EM are more or less the same as the initial ones. Only some of EM moved deeper to the core and some other moved near the tidal radius.

Initial positions of CBS and CBB correlate nicely with the core radius ($r_c$). CBS and CBB before the core collapse were created mainly inside the core radius in the dynamical interactions with wide binaries. During the core collapse and afterwards, they were created mainly in the core, but some of them also outside the core up to the half-mass radius. CBS and CBB are very often inside binaries, so they are heavier than EM or EMT and sink to the center of the cluster due to mass segregation. Some of them, because of the dynamical interactions, are ejected to the halo of the star cluster. Thus, their final positions is far outside the half-mass radius.

\subsubsection{Bimodal distribution}

\begin{figure*}
	\centering
	\includegraphics[width=\picWidth,angle=270]{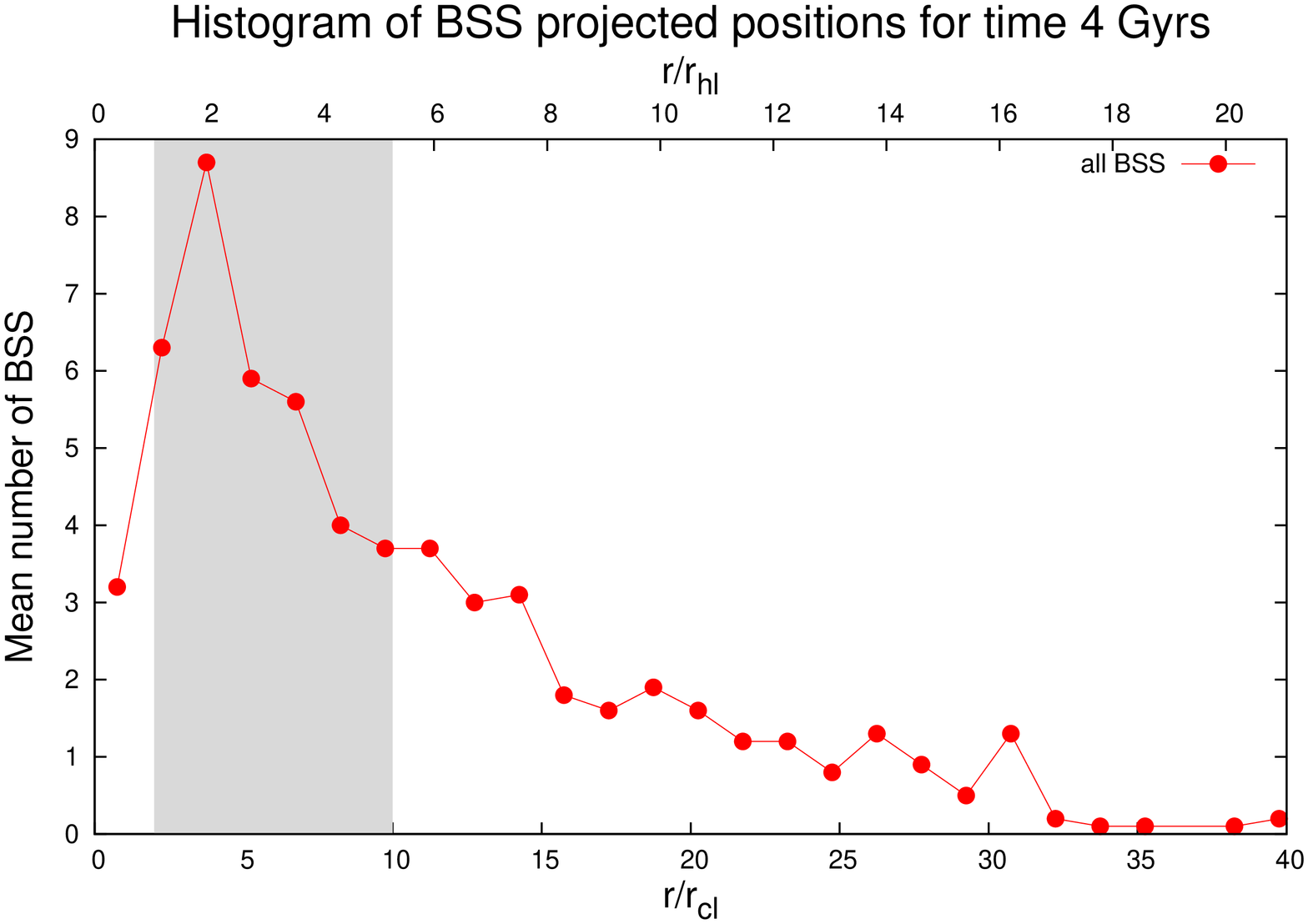}
	\includegraphics[width=\picWidth,angle=270]{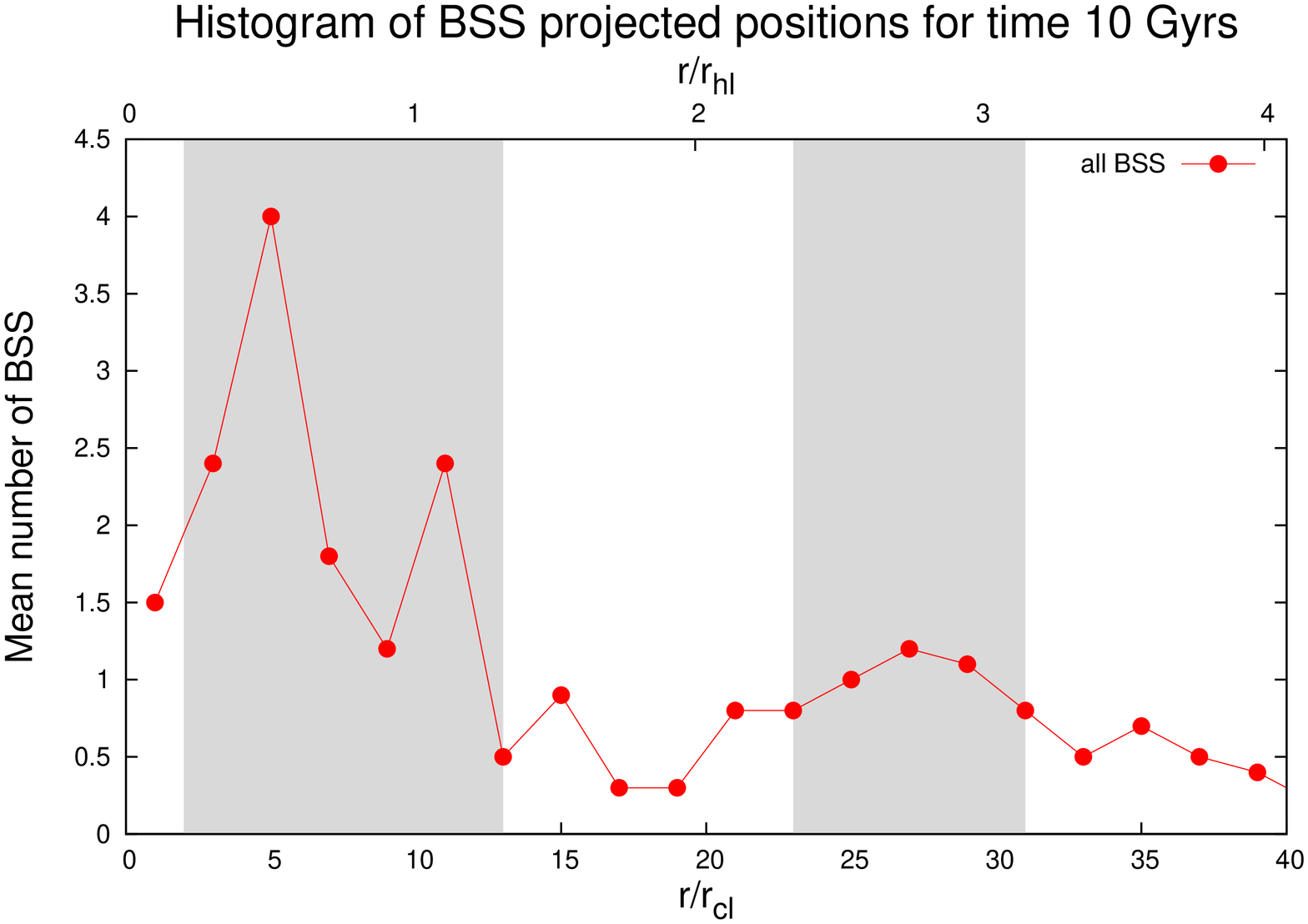}
	\caption{Histograms with BSS projected distances divided by $r_{cl}$ at the time 4~Gyrs, and time 10~Gyrs, for all BSS. Bimodality in the BSS distribution is not visible yet for the time 4~Gyrs, but a weak bimodality is visible for the time 10~Gyrs. Both radii, $r_{cl}$ and $r_{hl}$, are observational radii.}
	\label{pic:BSSBimodality}
\end{figure*}

Bimodal distribution of BSS was found in many star clusters (\citet{Ferraro1993AJ....106.2324F, 1997A&A...324..915F, Zaggia1997A&A...327.1004Z}, and other references in Sec.~\ref{sec:Introduction}). Projected distances of BSS for our test simulation are shown in the first panel in Fig.~\ref{pic:BSSBimodality} for the time 4~Gyrs, and in the second panel in Fig.~\ref{pic:BSSBimodality} for the time 10~Gyrs. Because the number of BSS is rather small, in order to decrease the statistical fluctuations, the projected distance was calculated 10 times for each BSS. In Fig.~\ref{pic:BSSBimodality} there are given mean numbers of BSS depending on the distance from the center. On the bottom label of the X-axis there is the distance relative to the core radius, and on the top label there is the distance relative to the half-light radius. Both, the core radius and the half-light radius, are observational ones (see definitions in Sec.~\ref{sec:InitialModel}). 

In the first panel of Fig.~\ref{pic:BSSBimodality} (time 4~Gyrs) one can see that the bimodality of BSS distribution is not present. The majority of BSS is located near 2 half-light radii. The BSS distribution continuously drops with the distance from the center of the star cluster. It seems that a weak bimodality becomes visible after the core collapse (the second panel of Fig.~\ref{pic:BSSBimodality}). One can see that the first high peak of the number of BSS is about 5~$r_{cl}$ (0.5 $r_{hl}$), a well defined minimum about 18~$r_{cl}$ (1.8~$r_{hl}$), and the second peak in the number of BSS is in the outskirts about 2.8~$r_{hl}$. In \citet{Ferraro1993AJ....106.2324F, 1997A&A...324..915F} the plot with the bimodal distribution for M3 has wider bins in the histogram for the second peak value. However, in our plot the width of bins is the same for all distances, thus in our model the second peak is less visible than in these cited papers. 

In our next paper we plan to investigate the bimodality for various star cluster properties and we will check for which star clusters and when the bimodality starts to appear. Perhaps we will find that the bimodality is present only in star clusters after the core collapse? It would be a great tool to probe dynamical stages of stars clusters.

\subsubsection{BSS lifetimes}
\label{sec:LifetimesOfBlueStragglers}

\begin{figure*}
	\centering  
	\includegraphics[width=\picWidth,angle=270]{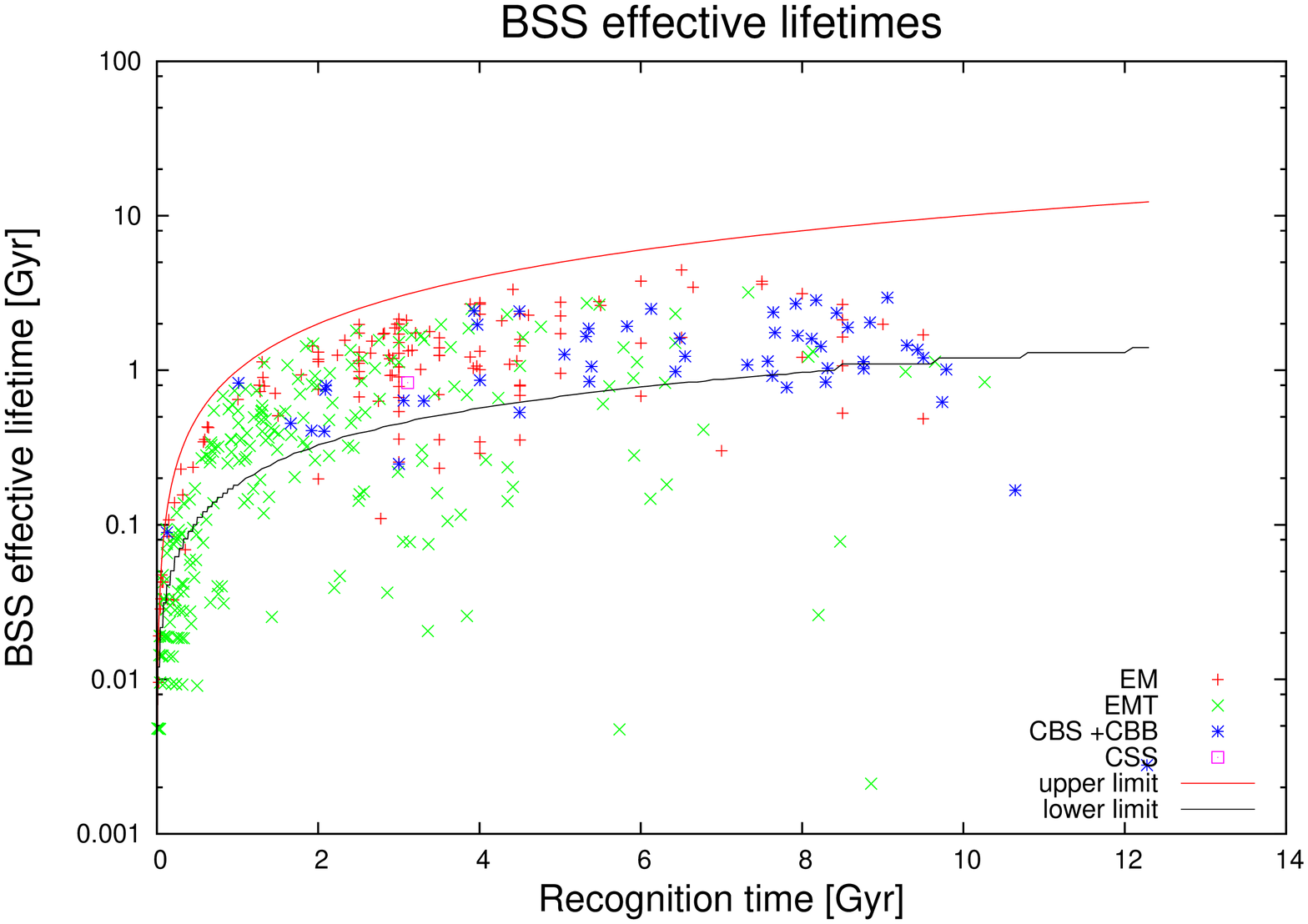}  
	\includegraphics[width=\picWidth,angle=270]{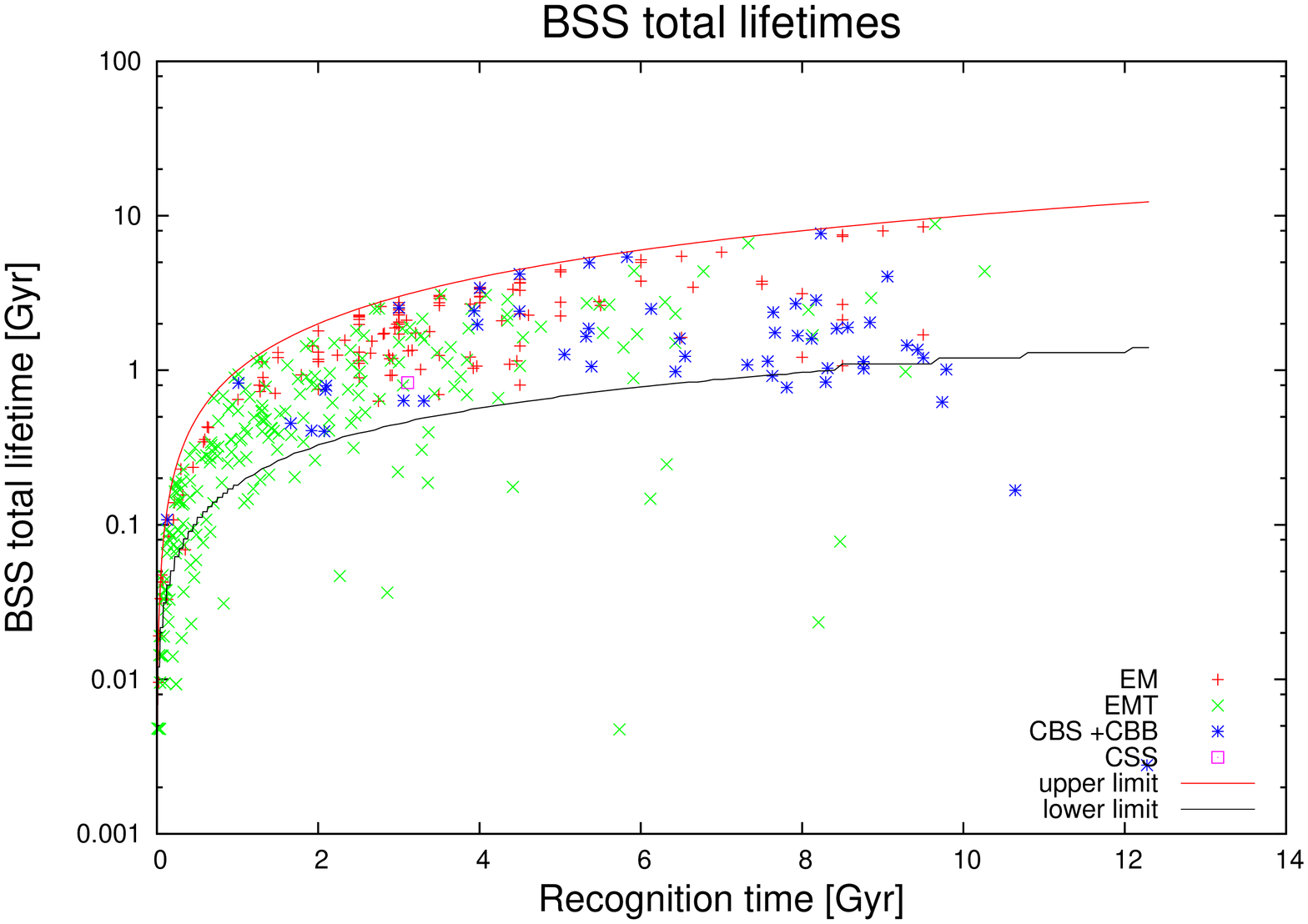}   
	\caption{Left: the effective lifetimes of BSS [Gyrs] at the recognition time [Gyrs] for different channels of formation, calculated from the time range between the recognition time and the termination time. Right: the total lifetimes of BSS [Gyrs] at the recognition time [Gyrs] for different channels of formation, calculated from the creation time and the termination time. These two panels show that for many BSS there is a significant delay for a star before it can be recognized as a BSS (i.e. it exceeds the turn-off mass). The upper limit is the expected lifetime of a main sequence star with the mass equal to the turn-off mass, and the lower limit is the expected lifetime of a main sequence star with the mass equal to twice the turn-off mass.
	}       
	\label{pic:Lifetimes}
\end{figure*}

Fig.~\ref{pic:Lifetimes} shows the effective (left panel) and the total (right panel) lifetimes of BSS for different channels of formation. Just to remind, the effective lifetime is calculated as a difference between the recognition time and the termination time. The total lifetime is calculated from the creation time and the termination time. 
In Fig.~\ref{pic:Lifetimes} there are also presented two boundary lines: the upper and the lower limits. The upper limit is the expected lifetime of a star on the main sequence with one turn-off mass, and the lower limit is the expected lifetime of a star on the main sequence with two times the turn-off mass. These boundaries show what are the theoretical minimum and maximum lifetimes for a BSS which had a one merger event before becoming a BSS. One can see that in the first panel of Fig.~\ref{pic:Lifetimes} there are many BSS, from different channels of formation, for which lifetimes are significantly smaller than the lower limit. However, in the second panel of the Fig.~\ref{pic:Lifetimes} there are only several such cases which have still smaller lifetimes than the lower limit (details are given later in this section).

All BSS created in the first 1 Gyr, as expected, are short living BSS because these are the heaviest stars. In the beginning of the star cluster evolution there are many massive stars in binaries which drive the EMT channel to produce BSS through mass transfers. After a few hundreds Myrs, when the most massive stars leave the main sequence, the number of EMT starts to decrease and BSS are created more efficiently in other channels. Lifetimes of BSS produced after 1 Gyr are significantly longer than for EMT created earlier. The reason is that BSS created after the first Gyr are not as massive as in the beginning of the simulation. Their evolution is much slower and the BSS phase can last much longer.

\begin{figure}
	\includegraphics[width=\picWidth,angle=270]{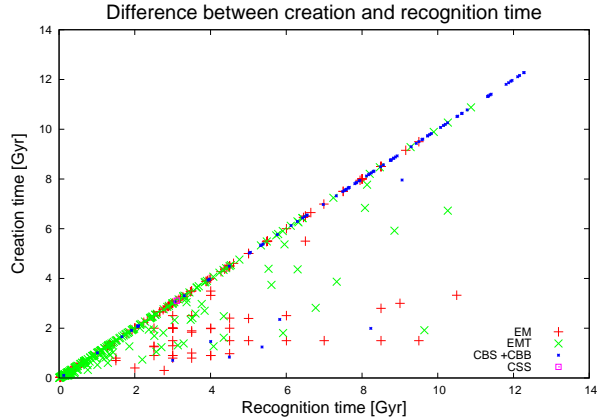}     
	\caption{The recognition time of BSS [Gyr] vs. the creation time [Gyr] for different channels of formation. The plot shows what is the difference in time between an event which really created BSS and the actual recognition of a BSS.}
	\label{pic:LastMergerTime}
\end{figure}

Fig.~\ref{pic:Lifetimes} shows how the BSS phase can be delayed before the actual detection by observations. The BSS last merger or the last mass transfer can happen even several Gyrs before the turn-off mass will decrease enough (see Fig.~\ref{pic:LastMergerTime}), so that star can be recognized as a BSS. Total lifetimes take into account delayed detection. In this case many of BSS moved with their lifetimes between the upper and lower limits. Additionally, one can see in the second panel of Fig.~\ref{pic:Lifetimes}, that many of BSS with corrected lifetimes also moved closer to the upper limit (red line).

In the second panel of Fig.~\ref{pic:Lifetimes} there are about 30 more EMT BSS (mainly for T $<$ 1~Gyrs) which have still total lifetimes significantly shorter than the lower limit, even taking into account a delayed detection. It was unexpected to have BSS created at later stages of the star cluster evolution with such short lifetimes (only $\sim 10$ Myrs). Some of these stars exceeded the turn-off mass by just a little and started a very short BSS phase. After just few Myrs (up to 100 Myrs) they evolved from MS to the red giant phase, ending the BSS phase. In a few other cases, instead of the red giant phase, there was a merger between the binary components which terminated the BSS phase (the resulting star was not a MS star). There are also two CBS/CBB with very short total lifetimes. For one of them, the BSS left MS very fast. For the second one, there occurred another merger event while the star was still BSS, but after that the star was not BSS any longer.  

Fig.~\ref{pic:LastMergerTime} shows a relation between the recognition time of BSS and the creation time. Every point which is below the imaginary straight line $x = y$ denotes BSS for which a merger or a mass transfer event occurred before the star was detected as a BSS. One can see that many BSS from EM and EMT channels, and some BSS from CBS and CBB channels had mergers even several Gyrs before they became BSS. Mergers or mass transfers occurred for these stars earlier, but masses were too small, so they did not exceeded the turn-off mass and they had to wait for a detection as a BSS. This effect was not expected in common scenarios for creation BSS, and it was rather assumed, that mergers between stars create BSS immediately. This plot shows that delayed detection of dormant BSS is significant. For all 149 EM BSS there were 45 dormant EM BSS (30\%), for all 231 EMT BSS there were 60 dormant EMT (26\%) and for all 95 CBS and CBB there were only 7 dormant CBS, CBB BSS (7\%). For the total 476 BSS there were overall 112 dormant BSS which is 24\%.

\begin{figure}
	\includegraphics[width=\picWidth,angle=270]{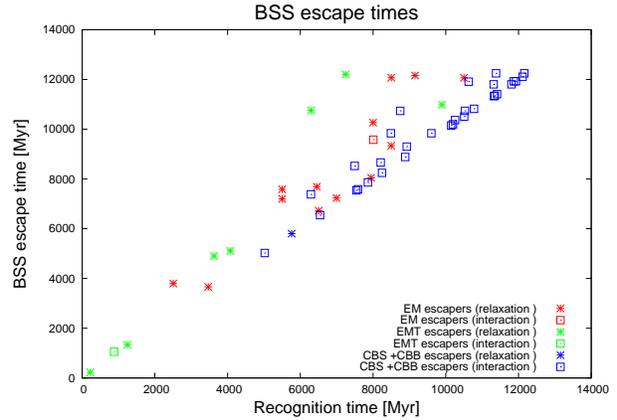}    
	\caption{The recognition time [Gyr] vs. the escape times of BSS [Gyr] for different channels of formation. In the plot there are only BSS which escaped as BSS. For each channel of formation there are two types of points showing the cause of the escape: relaxation or dynamical interaction.}
	\label{pic:Escapers}
\end{figure}

Fig.~\ref{pic:Escapers} shows the escape time of BSS depending on the recognition time. It shows the escape times for the main channels of formation: EM, EMT and CBS$+$CBB. In the figure there are only BSS which escaped from star cluster as BSS. All other BSS which first stopped being BSS, and then escaped from the star cluster, are not taken into account in this plot. Also, escaped BSS in Fig.~\ref{pic:Escapers} are divided into two groups, according to different reasons of the escape: due to energetic dynamical interaction (empty squares in the plot) and due to the relaxation process (stars in the plot). One can see that the number of BSS escapers increases during later stages of the cluster evolution, when the core starts to collapse ($\sim 6$ Gyrs). The number of escapers increases most significantly after the core collapse ($> 8$ Gyrs). Additionally, there are significantly more BSS escapers from CBS$+$CBB channels, for which the time between the creation and the escape time is small (fast escapers). This is caused by the fact that BSS created due to dynamical interactions gain significant kinetic energy, which allows them to leave the star cluster faster. All BSS from CBS and CBB channels, except just one, escaped from the system due to the dynamical interactions (blue squares). In contrast, for EM and EMT BSS the process responsible for the escape from the system is the relaxation. There is just one EM case and one EMT case for which the dynamical interactions were the reasons of the escape. Relaxation
do not give BSS often such a large additional kinetic energy as the dynamical interactions. Thus, there are in general less EM and EMT which can leave the system and they need significantly more time for the escape (slow escapers, see Fig.~\ref{pic:Escapers}).

From the overall 476 BSS created in the simulation, 62 escaped as BSS ($\sim 13\%$). From the total 231 EMT BSS created during the whole simulation, only 8 of them escaped as BSS ($\sim 3\%$), and from the total 149 EM BSS, only 14 escaped as BSS ($\sim 9\%$). However, BSS escapers are more important for CBS and CBB channels because for the total 95 of them, 40 escaped as BSS ($\sim 43\%$). CBS and CBB escapers become even more important after the core collapse ($>$ 8~Gyrs), when there were created 50 CBS and CBB and 30 of them escaped as BSS ($60\%$). 
For our test model on average over 10\% of BSS escape from the cluster, so it seems that escape BSS process could be important. Thus, one can try to search for them in tidal tails of post-collapse stars clusters (especially slow escapers).

\subsubsection{BSS in binaries}

\begin{figure}
	\includegraphics[width=\picWidth,angle=270]{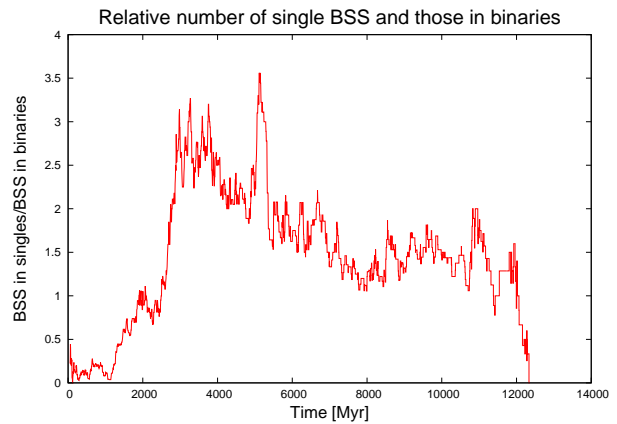}
	\caption{The ratio between the number of BSS as single stars and the number of BSS inside binaries as a function of time}
	\label{pic:BSS_relative_single_binary}
\end{figure}

The ratio between the number of single BSS and the number of BSS in binaries as a function of time is shown in the Fig.~\ref{pic:BSS_relative_single_binary}. In other words, plot shows how many times there is more BSS as single stars than BSS in binaries. 

The ratio starts favoring BSS in binaries. The main BSS channel responsible for that is EMT. The mass transfer creates BSS in primordial, massive binaries, but the EMT channel dominates only for the first $\sim 2$~Gyrs of the star cluster evolution. After that time EM start to play a bigger role, as single BSS. The number of BSS from EM channel continuously increases, while EMT BSS continuously decreases, until there is 2-3 times more BSS as single stars than in binaries. The shape of the plot, especially its peaks, corresponds nicely to the peak of the EM channel (see Fig.~\ref{pic:Time_vs_BSSInitialType}). When EM channel starts to drop after $\sim 4$~Gyrs, BSS in singles to BSS in binaries ratio starts to drop as well. The ratio drops to $\sim 1.5$. It is interesting that it stays around this value for the next several Gyrs. Just at the end of the star cluster evolution ($> 12$~Gyrs) the ratio flips again, and there are more BSS in binaries again. This is caused by the fact that at the end of the star cluster evolution there are more binaries left, because of their heavier mass.

\section{Discussion and summary}
\label{sec:Summary}

We described here an improved code for simulations of star clusters, called MOCCA, which stands for MOnte Carlo Cluster simulAtor. It combines two very distinctive approaches for the star clusters simulations: the Monte Carlo method and the \textit{Fewbody} -- a direct integration method. The \textit{Fewbody} is used to perform small N-body scattering experiments between binaries and single stars and between two binaries. This gave the current version of the code very needed features. From now on, one can follow the evolution of exotic objects formed in complicated dynamical interactions, like BSS. The MOCCA code was used here to investigate a population of different channels of formation of BSS in star cluster environment with 100k initial objects (80k single stars and 20k primordial binaries). This model was chosen as a test model to check the ability of the code to follow the evolution of BSS. BSS were divided in this paper into two major categories: evolutionary and dynamical. To the evolutionary category of BSS we include evolutionary mergers (EM), evolutionary mass transfers (EMT) and evolutionary dissolutions (ED). EM is created when two stars from a binary merge as a result of the binary stellar evolution (without dynamical interaction with other stars in the system). EMT is a BSS which is created due to a mass transfer, which allows the star to exceed the turn-off mass. ED is a BSS which is created when a binary dissolves during the supernova explosion. To the dynamical category of BSS we include all BSS which were created as a result of the dynamical interactions between two single stars (CSS), or between single star and binary (CBS), or between two binaries (CBB). There are also other types of BSS introduced in this paper, but they rather represent changes of types of BSS connected with the dynamical interactions, not new channels of formation.

Some conclusions in this paper are not general, because we discussed here only one test model. In the next papers we plan to perform many more simulations with various initial parameters (both global and for binaries properties) and we will try to study in details the channels of formation of BSS and conclusions presented here.

For EMT channel we noticed groups of BSS which had different physical processes responsible for their creation. The first group consists of EMT BSS which are created through a Roche lobe overflow in compact binaries -- short period EMT.
In this scenario one of the stars in the binary starts to leave the main sequence. 
It increases its radius, and at some point, a mass starts to flow to the other star, which later on becomes the EMT BSS. The second group of EMT BSS is formed in the long period binaries in a different way. Here, a donor star at some point starts to leave the main sequence, enters the giant branch and ejects some mass through stellar winds. 
The star, which later on becomes BSS, gains a mass (about few $\sim 0.01 \mathrm{M_{\odot}}$), which is not enough to become a BSS. The donor star then evolves to short AGB phase, ejects an envelope. This time, the future BSS, increases the mass by about $< 0.1 \mathrm{M_{\odot}}$. The donor star quickly becomes a white dwarf, but the star which gains mass, in many cases still does not exceeds the turn-off mass. This is why these types of EMT are in majority dormant BSS. It has to take some time, even several Gyrs, until the star exceeds the turn-off mass and becomes a BSS. Thus, the most distinct difference, in comparison to short period EMT, is that long period EMT gain only small additional mass, wait until they exceed the turn-off mass, and in general are short living BSS because they exceed the turn-off mass only slightly. EMT are created from around the core up to far outside of the half-mass radius. Final EMT positions are a little bit more scattered in comparison to the initial ones. Some of EMT stayed near and around the core and some moved outside the half-mass radius, closer to the tidal radius.

Mass ratios of the long period EMT at the recognition time create a distinct trend (see the bottom-right panel in Fig.~\ref{pic:EMTSubgroups}). The trend is very well reproduced with the Eq.~\ref{eq:LongEMTQ}. The masses of BSS are just slightly larger than the turn-off mass, thus in the nominator of Eq.~\ref{eq:LongEMTQ} there is $M_{turn-off}$. WDs are companions in the long period EMT, so in the denominator there are calculated masses of WDs based on \citet[Tab. 1]{Chernoff1990ApJ...351..121C}. The mass ratios for long period EMT have a narrow range and predictable values through the entire simulation. \citet{Geller2011Natur.478..356G} reported that BSS in long period binaries in an old (7 Gyr) open cluster, NGC~188, have companions with masses of about half of the solar mass (surprisingly narrow mass distribution). This rules out a collisional origin for these long period BSS, because otherwise, for collision hypothesis there would be significantly more companions with higher masses. It is consistent with a mass transfer origin for the long-period blue straggler binaries in NGC 188, in which the companions would be white dwarfs of about half of a solar mass \citep{Geller2011Natur.478..356G}. This finding is very consistent with our work. We find for long period EMT that their mass ratios are in fact narrow. Mass ratios decrease according to the Eq.~\ref{eq:LongEMTQ}. There are indeed WDs in long period EMT with masses of about $0.5 \mathrm{M_{\odot}}$ at the time 7~Gyrs.  

EM are created from compact binaries with eccentricities close to zero. For EM BSS we noticed two different formation processes as well. The first one creates EM through the Roche lobe overflow during a semi-detached phase of a binary. The second one creates EM BSS from a magnetic braking. Although both groups overlaps each other during the simulation, EM created from the Roche lobe overflow are created mainly in the beginning of the simulation, and EM created from magnetic braking are formed mainly after several Gyrs.
All EM were created around the core and near the half-mass radius. The final positions of EM were more or less the same as initial ones with the exception that the final positions are a little bit more scattered in the cluster: from the core up to even the tidal radius -- similarly to EMT.

Collisional BSS (CBS and CBB) are created in strong dynamical interactions. 
Before the core starts to collapse ($< 6$~Gyrs), CBS and CBB are created in the core from binaries with mostly large orbital periods (about $10^4$ days).
During and after the core collapse, when the core is getting denser, the number of CBS and CBB increases. Also, they are being created from more compact binaries which have orbital periods of about a few days. BSS from dynamical channels are formed mainly in the core or close to it. It is actually not surprising because in order to increase the chance of a collision in the dynamical interactions, a dense environment is needed. Some of BSS in the dynamical interactions gained more additional kinetic energy, extended their orbits, moved near the tidal radius and beyond. Many of them escaped from the star cluster as BSS.  

EMT BSS are the most active in the beginning of the star cluster evolution, its population continuously drops but nevertheless stays active during the whole cluster evolution. The population of EM BSS is not significant in the beginning. Its peak value in our simulation is about $\sim$ 3~Gyrs, when both scenarios work most efficiently for EM -- the Roche lobe overflow and magnetic braking. Additionally, the magnetic braking starts to work for both components in binaries (the turn-off mass drops at that time to about 1.25~$\mathrm{M_{\odot}}$, see \citet{Hurley2002MNRAS.329..897H} for details). CBS$+$CBB channels starts to play a significant role in the overall population of BSS in the post collapse phase. According to \citet{1992AJ....104.1831F} for less dense GCs, BSS could form as evolutionary mergers, in contrary to dense GCs, for which BSS could form from dynamical interactions. Our test simulation is consistent with it. Before the core collapse, there are more evolutionary mergers and the dynamical interactions become important after the core collapse. Nevertheless, we have BSS from different channels during the whole simulation. Moreover, we think that the number of BSS from evolutionary channels (EM, EMT) depend rather on initial binary properties. We expect to have less binaries for EMT and EM channels for star clusters where there are less compact primordial binaries. For simulations with larger initial concentrations we also expect to have more BSS from CBS, CBB channels and maybe also from CSS. Nevertheless, we expect to have BSS from all channels of formation for various densities of the star clusters. The only exception could be for the CSS channel. In our test model we had only one BSS created by the collision of two field stars, and even after the core collapse, the number of CSS BSS did not increase. Perhaps for star clusters with higher densities CSS will increase its overall significance. BSS channels of formation work simultaneously in different radial parts \citep{1997A&A...324..915F, 2006MNRAS.373..361M}. Indeed, based on our test model, one can see that EM and EMT are created more or less in the same radial distances from the center. Only BSS from CBS and CBB channels are created mainly inside and around the core, so in deeper regions than EM and EMT.

Number of BSS does not correlate with the predicted collision rates \citep{2004ApJ...604L.109P, 2008ApJ...678..564L, 2008IAUS..246..331L}, which is one of the reasons why \citet{Knigge2009Natur.457..288K} favor the mass transfer mechanism as a more important scenario for creation of BSS. Our test simulation shows that there are indeed many EMT created (49\% of all BSS are EMT) and that this channel is active for the whole simulation. However, the number of EMT drops from the beginning and at some point the number of BSS created by mergers and collisions (EM, CBS, CBB) outnumbers the EMT BSS. Moreover, we think that the population of evolutionary BSS depends mainly on initial binary properties and thus, for different initial conditions EMT could be indeed the most important channel of formation. A smaller number of primordial compact binaries could create less EM. A stronger dependence on the initial binary properties, rather than star cluster global parameters, like concentration, are also consistent with the calculated specific frequencies from \citet{2003ApJ...588..464F}. They found that the largest BSS specific frequencies are observed for the star clusters with the lowest central density (NGC~288) and the highest central density (M80). It suggests, that the number of BSS depends rather on the initial binary properties, instead of the global parameter -- the central density. \citet{2008A&A...481..701S} found the linear correlation between the number of BSS and the binary fraction, which in turn depends mainly on the initial binary properties of a star cluster. Thus, it is very important to find an observational distinction between the mass transfer BSS and the collisional ones (first attempts are already done by \citet{2009RMxAC..37...62F}). \citet{2008A&A...481..701S} suggest that the strong correlation between the number of BSS and the binary fraction is a result of formation channel of BSS as the unperturbed evolution of the primordial binaries. They did not found correlations with the central density, the concentration, the stellar collision rate, and the half-mass relaxation time of star clusters. However, later \citet{Knigge2009Natur.457..288K} claimed that if one restricts a sample of star clusters to these with dense cores, a significant correlation appears between the number of BSS and the collision rate. We plan to check both, different initial binary properties and different initial global properties, and perform many simulations to study how much the population of BSS depends on initial binary properties, collision rates and concentrations.

BSS created as EM or EMT very rarely change their types. Only in a few cases EM got into a binary in some dynamical interactions and some EMT were disrupted or collided with other stars during the dynamical interactions. EM do not change types most likely because they are single stars, and the probabilities of the dynamical interactions for them is lower than for EMT. Additionally, EM are not the most massive objects and the number of EM is small in comparison to the number of e.g. WDs at the later stages of the star cluster evolution. EMT do not change types, probably because of the fact, that most of them are hard enough to survive dynamical interactions. Additionally, the number of EMT is even smaller than the number of EM, when the dynamical interactions become important ($> 8$~Gyrs). One of the subgroups of EMT are long period BSS which should favor the higher probabilities of the dynamical interactions. However, long period EMT are short living BSS and thus, they have small chances to be changed by the dynamical interactions.
As expected, CBS and CBB channels change their types quite often. They are disrupted (if BSS was in a binary) or exchanged with other stars in dynamical interactions. In many cases, CBS and CBB, stay in binaries or get into heavier binaries quickly and thus, they have larger masses in general. Additionally, CBS and CBB are created mainly in the core or around it. Due to the mass segregation they sink to the center of a cluster. All these facts are the reasons why the probabilities of the dynamical interactions for CBS and CBB are in general higher than for EM or EMT and thus they can change types more frequently.

Many of EM and EMT BSS, which are created after the core collapse, had some strong dynamical interactions before they became BSS. The strong dynamical interactions means that the semi-major axis or the eccentricity was changed by some dynamical interaction at least by 10\% (arbitrary chosen value for our test simulation). It suggests that such EM and EMT could be created by possible induced mass transfers or possible induced mergers. In other words, the creation of EM and EMT could be triggered or made possible by the dynamical interactions. However, for the simulation without the \textit{Fewbody}, and for simple population synthesis, the overall number of EM and EMT were almost the same (see Sec.~\ref{sec:ComparionOldVersion}, Fig.~\ref{pic:InitialBssNoFB}). This, in turn, supports the theory that the dynamical interactions are not important for the populations of EM and EMT. On the other hand, for a simulation with higher $r_{pmax}$ (higher impact parameters) in the MOCCA code, there are significantly more fly-by interactions for binaries (for details see the next paper in the series \citet[Fig.~7]{Giersz2011arXiv1112.6246G}). In such a case there were also created many more BSS. This, in turn, suggests that the dynamical interactions could have some influence on the creation of BSS. We plan to check in the next papers whether the dynamical interactions have indeed some implications for the population of EM or EMT. Maybe the star clusters with higher concentrations, or for which dynamical interactions are more important in overall evolution, will have larger population of EM and EMT. 

A bimodal distribution of BSS is present in many star clusters. It was discovered by \citet{Ferraro1993AJ....106.2324F, 1997A&A...324..915F} for M3 and by \citet{Zaggia1997A&A...327.1004Z} for M55. Later, the bimodality was observed for many other clusters and for a few of them it was not present, e.g. for NGC~2419 \citep{2008ApJ...677.1069D,Contreras2012ApJ...748...91C}. We were also able to see weak signs of the bimodal distribution of BSS in our test simulation. The bimodal distribution of BSS was explained by \citet{1997A&A...324..915F} as a result of different processes forming BSS in different parts of the cluster -- mass transfers for the outer BSS and stellar collisions for BSS in the center. Later, \citet{2004ApJ...605L..29M, 2006MNRAS.373..361M} and \citet{2007ApJ...663..267L}, showed that the bimodal distribution in the cluster cannot be explained only by the collisional scenario, when BSS are created in the center of the cluster and some of them are ejected to the outer parts of the system. This process is believed to be not efficient enough and $\sim 20-40\%$ of BSS have to be created in the peripherals in order to get the required number of BSS in the star clusters. It is believed that in the outer parts of the star cluster binaries can start the mass transfer in isolation without suffering from the energetic dynamical interactions with field stars. The last explanation is very consistent with our test simulation. BSS from EM and EMT channels are created from the core up to the outside of the half-mass radius, and CBS+CBB are created mainly inside and around the core (see Sec.~\ref{sec:BssPositions}). Some of CBS and CBB are indeed ejected to more extended orbits. In total, the number of BSS outside the half-mass radius consist of all channels (EM, EMT and CBS+CBB). Moreover, the bimodality of BSS is not present always in our test simulation, but it seems to form after the core collapse and becomes visible after that. If this could be confirmed in our next planned simulations with different initial conditions, the bimodality could be a tracer of a dynamical status of a cluster.

We investigated lifetimes of BSS and noticed that for some BSS there is a significant delay between the creation time (the time of the last merger or the last mass transfer) and the recognition time (the time when a star actually exceeded the turn-off mass). 
For some BSS this delay can last even several Gyrs. Such a delay was unexpected to find. BSS, for which such a delay exists, we call dormant BSS. The number of dormant EM and EMT is significant. For all 149 EM BSS there were 45 dormant EM BSS (30\%), for all 231 EMT BSS there were 60 dormant EMT (26\%) and for all 95 CBS and CBB only 7 were dormant (7\%). For the total 476 BSS there were overall 112 dormant BSS which is 24\%.

There is a number of BSS which escaped from the star cluster. For EM and EMT, this process seems to be not important because only respectively 3\% and 9\% escaped as BSS. However, for CBS and CBB channels for the whole simulation, 40 BSS (from the total 95) escaped as BSS, which gives 43\% efficiency. The number of CBS and CBB escapers increase even more if one narrow the time to the post collapse. After the core collapse, 60\% of CBS and CBB escaped as BSS from the star cluster. There are known BSS which are found in the halo and in the bulge of the Galaxy \citep{Bragaglia2005IAUS..228..243B, Fuhrmann2011MNRAS.416..391F, Clarkson2011ApJ...735...37C}, and there are also known fast moving BSS \citep{Tillich2010A&A...517A..36T}. These BSS were probably created from CBS or CBB channels, because from the dynamical interactions these stars could get high escape velocities. Our test model seems to show that CBS and CBB escapers could be important. 
Also, it seems that for BSS escapers channel of formation determines what is the cause that BSS leaves the system. If EM and EMT are ejected from the star cluster, it is because of the relaxation processes. CBS and CBB, in the contrary, leave the system because of the dynamical interactions -- they gain additional kinetic energy. CBS and CSS are fast escapers, whereas EM and EMT are slow ones. For EM and EMT there is significant delay between the recognition time and the escape time (see Fig.~\ref{pic:Escapers}).  We plan to check in the next paper how significant BSS escapers will be for all channels of formation for different initial parameters.

\subsection{Future plans}

In the next papers we plan to perform a systematic study for different binary properties, different IMF, concentrations, metallicities and many other star cluster parameters. We plan to check how it can influence the overall population of BSS or the population of the different channels of formation. We plan to search for correlations between results of the simulations and observational data.

The \textit{Fewbody} code allows to deal with any kind of hierarchies like triples, quadruples and higher hierarchies. However, currently the MOCCA code is unable to deal with such type of objects. It works only with single and binary stars. One of the significant new features would be to implement procedures to deal with the higher hierarchies.

Another major new feature would be parallelization of the source code. The usage of the \textit{Fewbody} slowed down the code. In order to regain, as much as possible, the previous performance of the MOCCA code, the next step is to parallelize the main bottlenecks of the code. Executing the dynamical interactions in parallel could give some speed up. A natural choice is to use OpenMP because the main procedures are just loops which sequentially execute specific functions.

\section{ACKNOWLEDGEMENTS}
We would like to thank J.~M.~Fregeau for giving us permission to use the \textit{Fewbody} code and his help with its usage. AH would like to thank Professor Douglas Heggie for his hospitality during the visit in Edinburgh and his comments about the draft version of the paper. Thanks also go to Professor Joanna Miko{\l }ajewska, Professor Micha{\l } R\'o\.zyczka, Professor Janusz Zi\'o\l kowski and Professor Janusz Ka\l u\.zny whose comments and suggestions improved this paper very much. 

The project was supported by Polish Ministry of Sciences and Higher Education through the grants 92/N-ASTROSIM/2008/0, N~N203~38036 and 
Polish National Science Center grant DEC-2011/01/N/ST9/06000.

\bibliographystyle{mn2e} 
\bibliography{2010_FewBody}

\bsp

\label{lastpage}

\end{document}